\pdfoutput=1
\documentclass[final,5p,times,twocolumn]{elsarticle}

\usepackage{graphicx}
\usepackage{amssymb}
\usepackage{amsmath}
\usepackage{enumitem}
\usepackage{array}
\usepackage{booktabs}
\usepackage{multirow}
\usepackage{ragged2e}
\usepackage{geometry}
\usepackage{multicol}
\usepackage{longtable}
\usepackage{adjustbox}
\usepackage{caption}
\usepackage{tabularx}
\usepackage{supertabular}
\usepackage{xcolor}
\usepackage{tikz}
\usepackage{forest}

\usepackage[colorlinks=true,linkcolor=blue,citecolor=blue,urlcolor=blue]{hyperref}

\definecolor{colorTit1}{RGB}{230, 210, 230}

\definecolor{colorSeg1}{RGB}{207, 222, 250}
\definecolor{colorSeg2}{RGB}{233, 240, 252}

\definecolor{colorDig1}{RGB}{207, 233, 213}
\definecolor{colorDig2}{RGB}{238, 247, 240}

\definecolor{colorGer1}{RGB}{250, 225, 220}
\definecolor{colorGer2}{RGB}{254, 235, 232}

\definecolor{colorFM}{RGB}{250, 245, 220}
\definecolor{colorFM2}{RGB}{254, 250, 240}

\usetikzlibrary{shadows.blur, shapes.multipart, fit, arrows.meta}

\newcolumntype{L}[1]{>{\RaggedRight\hspace{0pt}}p{#1}}

\author[1]{Xiaoling Luo}

\author[1]{Ruli Zheng}

\author[1]{Qiaojian Zheng}

\author[1]{Zibo Du}

\author[1]{Shuo Yang}

\author[1]{Meidan Ding}

\author[2]{Qihao Xu}

\author[3]{Chengliang Liu\corref{cor1}}
\cortext[cor1]{Corresponding author: Chengliang Liu, Linlin Shen;
  Email: liucl1996@163.com, llshen@szu.edu.cn}
  
\author[4]{Linlin Shen\corref{cor1}}

\address[1]{College of Computer Science and Software Engineering, Shenzhen University, Shenzhen, China}
\address[2]{Shenzhen Key Laboratory of Visual Object Detection and Recognition, Harbin Institute of Technology, Shenzhen, 518055, China}
\address[3]{Laboratory for Artificial Intelligence in Design, Hong Kong}
\address[4]{School of Artificial Intelligence, Shenzhen University, Shenzhen, China}

\journal{Journal}

\begin{document}

\begin{frontmatter}

\title{A Survey of Multimodal Ophthalmic Diagnostics: From Task-Specific Approaches to Foundational Models}




\begin{abstract}
Visual impairment represents a major global health challenge, with multimodal imaging providing complementary information that is essential for accurate ophthalmic diagnosis. This comprehensive survey systematically reviews the latest advances in multimodal deep learning methods in ophthalmology up to the year 2025. The review focuses on two main categories: task-specific multimodal approaches and large-scale multimodal foundation models. Task-specific approaches are designed for particular clinical applications such as lesion detection, disease diagnosis, and image synthesis. These methods utilize a variety of imaging modalities including color fundus photography, optical coherence tomography, and angiography. On the other hand, foundation models combine sophisticated vision-language architectures and large language models pretrained on diverse ophthalmic datasets. These models enable robust cross-modal understanding, automated clinical report generation, and decision support. The survey critically examines important datasets, evaluation metrics, and methodological innovations including self-supervised learning, attention-based fusion, and contrastive alignment. It also discusses ongoing challenges such as variability in data, limited annotations, lack of interpretability, and issues with generalizability across different patient populations. Finally, the survey outlines promising future directions that emphasize the use of ultra-widefield imaging and reinforcement learning-based reasoning frameworks to create intelligent, interpretable, and clinically applicable AI systems for ophthalmology.
\end{abstract}


\end{frontmatter}

\section{Introduction}
\label{sec1}
In recent years, artificial intelligence (AI) has achieved significant advances, driven by continuous improvements in machine learning methods. Among these advances, deep learning has emerged as the most transformative branch of machine learning, delivering unprecedented breakthroughs in ophthalmic diagnosis and treatment \cite{intro1, intro2}. Also known as representation learning \cite{intro4}, deep learning has become the predominant approach within medical AI due to its superior predictive performance, robust feature extraction capabilities, and strong adaptability \cite{intro3}. With rapid progress in computer vision, deep learning techniques demonstrate natural compatibility with ophthalmic medical imaging. By extracting latent pathological features from multidimensional ophthalmic data, such as color fundus photography (CFP) and optical coherence tomography (OCT), deep learning has significantly enhanced diagnostic accuracy for major vision-threatening diseases, including diabetic retinopathy (DR), glaucoma, and age-related macular degeneration (AMD) \cite{intro5}. As medical big data rapidly expands and algorithmic models undergo continual optimization alongside swift advancements in computational hardware, deep learning demonstrates significant potential for clinical translation in ophthalmic disease screening, lesion segmentation, and prognosis prediction.

\begin{figure*}[t]
\centering
\includegraphics[width=1\textwidth]{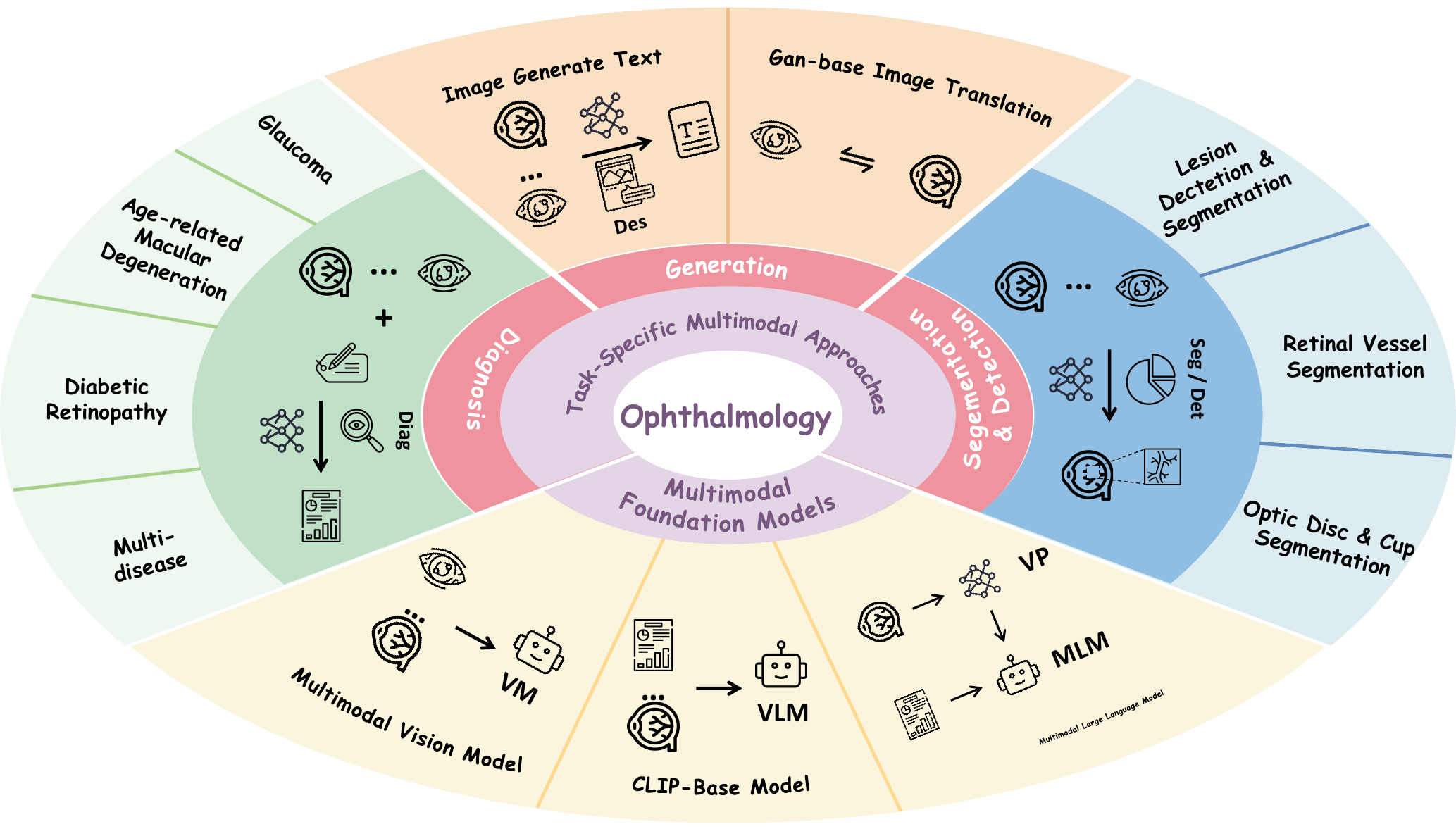}
\caption{Overview of multimodal model categorization in ophthalmology.}
\label{fig1}
\end{figure*}

In real-world clinical practice, ophthalmologists seldom rely on a single imaging modality for decision-making. Instead, they routinely integrate information from multiple imaging sources to obtain a more comprehensive view of retinal pathology. For instance, it is common to combine CFP, OCT, and OCT angiography (OCTA) to jointly evaluate retinal structural morphology, hemodynamics, and lesion functionality \cite{intro6}. Particularly in the context of early DR screening, incorporating modalities such as scanning laser ophthalmoscopy (SLO), OCT/A, and adaptive optics has been shown to enable the detection of subtle microcirculatory abnormalities and microleakage features, which are often difficult to capture using any single modality alone \cite{intro7}. By providing unprecedented accuracy, scalability, and efficiency, recent advances in multimodal deep learning are reshaping the paradigm of ophthalmic AI and are expected to ultimately enhance both patient outcomes and clinical workflows. Several studies have consistently demonstrated three key advantages of multimodal fusion over unimodal approaches: 1) enhanced diagnostic performance, with multimodal models outperforming unimodal models across accuracy, specificity, and AUC metrics \cite{intro8}; 2) improved model robustness, with studies showing that the integration of techniques such as data augmentation and transfer learning can substantially improve model stability and reliability \cite{intro9}; and 3) superior generalization ability and downstream task performance, as evidenced by a recent review of multimodal approaches \cite{intro10}.

Concurrently, large foundation models have emerged as a transformative force for multimodal learning, offering capabilities far beyond those of traditional architectures. Prominent examples such as CLIP, Flamingo, and GPT-4V have been pre-trained on massive and heterogeneous datasets encompassing images, text, and other modalities, allowing these models to generalize across diverse tasks with minimal supervision \cite{CLIP, Flamingo,gpt4}. In the field of ophthalmology, foundation models offer a unified framework that can integrate a wide range of clinical data, including fundus images, OCT scans, textual reports, and even genomic information, thereby enabling richer contextual understanding and improved diagnostic accuracy. Moreover, their ability to transfer knowledge across modalities makes them particularly valuable in resource-limited settings, where annotated medical data are scarce \cite{intro14}. As such, foundation models are poised to redefine the frontier of multimodal ophthalmic AI, bridging research silos and accelerating the deployment of intelligent systems in real-world clinical practice \cite{intro15}.

Despite these encouraging advances, the research landscape for multimodal deep learning in ophthalmic imaging remains highly fragmented, and no comprehensive review currently exists to systematically integrate methods, modalities, and clinical applications. Existing surveys have primarily focused on robustness and security challenges in medical AI \cite{intro9} and rarely address multimodal ophthalmic deep learning from a broad and unified perspective. To address this gap, the present survey provides a systematic and in-depth review of the state of multimodal deep learning in ophthalmic image analysis. This work mainly focuses on task-specific multimodal approaches and large-scale multimodal foundation models, which has a main organizational structure as shown in Figure~\ref{fig1}. We categorize existing methodologies, synthesize their core contributions, and highlight the key challenges that remain unresolved. Serving as a valuable reference for researchers, clinicians, and developers in the ophthalmic AI community, this work fosters cross-disciplinary collaboration and provides guidance for future innovations in this rapidly evolving field.

\section{Preliminaries}
\label{sec2}
Building on the motivations outlined in Section~\ref{sec1}, this section establishes the conceptual and methodological foundations necessary to systematically review multimodal deep learning in ophthalmology. We first clarify why this survey focuses on addressing the existing gaps in ophthalmic artificial intelligence research, where prior surveys have predominantly examined single‑modality approaches or narrow aspects such as robustness and security in medical AI \cite{pre1}. We then formally define the problem space and introduce the classification criteria that structure the remainder of this review, with particular emphasis on the distinction between task‑specific multimodal methods and large‑scale foundation models . Next, we delineate the scope of this review, including our literature retrieval strategy, inclusion and exclusion criteria, and the major ophthalmic imaging modalities considered as background knowledge (e.g., CFP, OCT/OCTA, FFA, and SLO) \cite{pre2, pre3}. Finally, we summarize related work to position our study in the context of existing reviews and highlight the unique contributions that differentiate this survey.
\subsection{Motivation}
Ophthalmic AI is undergoing a paradigm shift from unimodal analysis to multimodal intelligent decision-making. While deep learning has significantly improved diabetic retinopathy (DR) screening performance \cite{pre4}, real-world clinical applications reveal the limitations of single imaging modalities. For example, in the management of age-related macular degeneration (AMD), reliance on color fundus photography (CFP) alone can result in substantial underdetection of subretinal fluid (SRF) \cite{pre5}, while optical coherence tomography (OCT) alone fails to assess choroidal neovascular activity \cite{pre6}. This review provides the first systematic examination of multimodal deep learning in ophthalmology, aiming to guide the development of cross-modality diagnostic standards and next-generation AI systems.

\subsection{Problem Formulation}
In ophthalmology, the application of deep learning to multimodal data can be broadly classified into two primary paradigms: task-specific multimodal models and multimodal foundation models. This distinction highlights two fundamentally different strategies for managing the complexity of ophthalmic datasets. Task-specific models are tailored to specific clinical objectives and optimized for particular combinations of modalities to address clearly defined ophthalmic tasks. In contrast, foundation models leverage large-scale pretraining on heterogeneous data to achieve strong generalization across diverse sources and downstream applications. 

\subsubsection{Task-Specific Multimodal Models}
Specifically, task-specific multimodal models are typically designed as targeted solutions for well-defined clinical tasks. They exhibit a high degree of task specificity and are optimized for particular combinations of imaging and non-imaging modalities. These models are typically developed to achieve one or more of the following objectives: detection and segmentation, multimodal diagnostic, and image generation \& augmentation.

\textbf{Detection and segmentation}: In ophthalmic artificial intelligence, the detection and segmentation of anatomical structures and pathological lesions represent foundational tasks for disease screening, staging, and progression monitoring. These tasks aim to extract lesion regions, vascular networks, and optic disc/cup structures from multimodal imaging data. Three primary sub-directions are typically involved. Lesion segmentation focuses on extracting disease-specific pathological features, such as drusen, choroidal neovascularization (CNV), and macular edema in age-related macular degeneration (AMD), or microaneurysms and retinal hemorrhages in diabetic retinopathy (DR). Multimodal joint modeling significantly enhances localization accuracy. Vascular structure segmentation addresses the challenges posed by thin, low-contrast vessels, leveraging techniques such as ultra-widefield (UWF) and fluorescein angiography (FFA) fusion, as well as joint training with 3D and 2D data to improve vessel detection. Optic disc and cup segmentation, essential for glaucoma diagnosis and monitoring, benefits from multimodal integration and multi-task learning frameworks to improve delineation accuracy. With the increasing adoption of cross-modal attention mechanisms, multimodal fusion strategies, and semi-supervised learning paradigms, segmentation models are progressively overcoming the limitations of single-modality inputs, moving toward higher generalizability and clinical applicability.

\textbf{Multimodal Diagnostic}: Multimodal diagnostic models aim to integrate structural, functional, and clinical data to achieve early detection and fine-grained classification of ophthalmic diseases. Since certain pathologies may be inconspicuous or heterogeneous in a single modality, multimodal integration is essential. For instance, in early-stage AMD, lesion activity is often difficult to determine from color fundus photography (CFP) alone, but incorporating optical coherence tomography (OCT) or fundus autofluorescence (FAF) can significantly enhance the detection of reticular pseudodrusen (RPD) and CNV. In glaucoma, combining optic nerve structure (e.g., the cup-to-disc ratio) with functional visual field loss via OCT and perimetry data can reduce misdiagnosis caused by anatomical variability. DR presents a complex grading system, and its early-stage microvascular changes are often subtle; combining CFP with OCT angiography (OCTA) facilitates more accurate identification. Moreover, in real-world scenarios where multiple diseases often coexist (e.g., DR with concurrent AMD), single-task models struggle to capture the full spectrum of pathology. As a result, recent research trends are shifting toward joint multi-disease modeling and long-tailed sample enhancement strategies.

\textbf{Data Generation and Augmentation}: Ophthalmic image generation technologies such as modality translation and lesion description generation are increasingly employed to address challenges, including limited labeled data, high annotation costs, and suboptimal image quality. In modality translation, generative adversarial networks (GANs) have been utilized to convert standard CFP images into FFA, indocyanine green angiography (ICGA), or OCTA images, which enables non-invasive assessment of lesions. For example, generating multi-phase FFA from CFP facilitates monitoring of diabetic microvascular perfusion abnormalities. In image-to-text generation, models trained on aligned image-report pairs can automatically produce clinically relevant lesion descriptions. Recent advances that incorporate attention mechanisms, keyword guidance, and medical knowledge graphs have greatly enhanced the accuracy and logical consistency of the generated text. Furthermore, these generative techniques support downstream model training through data augmentation and pseudo-labeling, especially in scenarios with limited resources or across different domains. Overall, these innovations not only expand the availability of training data but also provide new possibilities for decision-support systems in clinical ophthalmology.

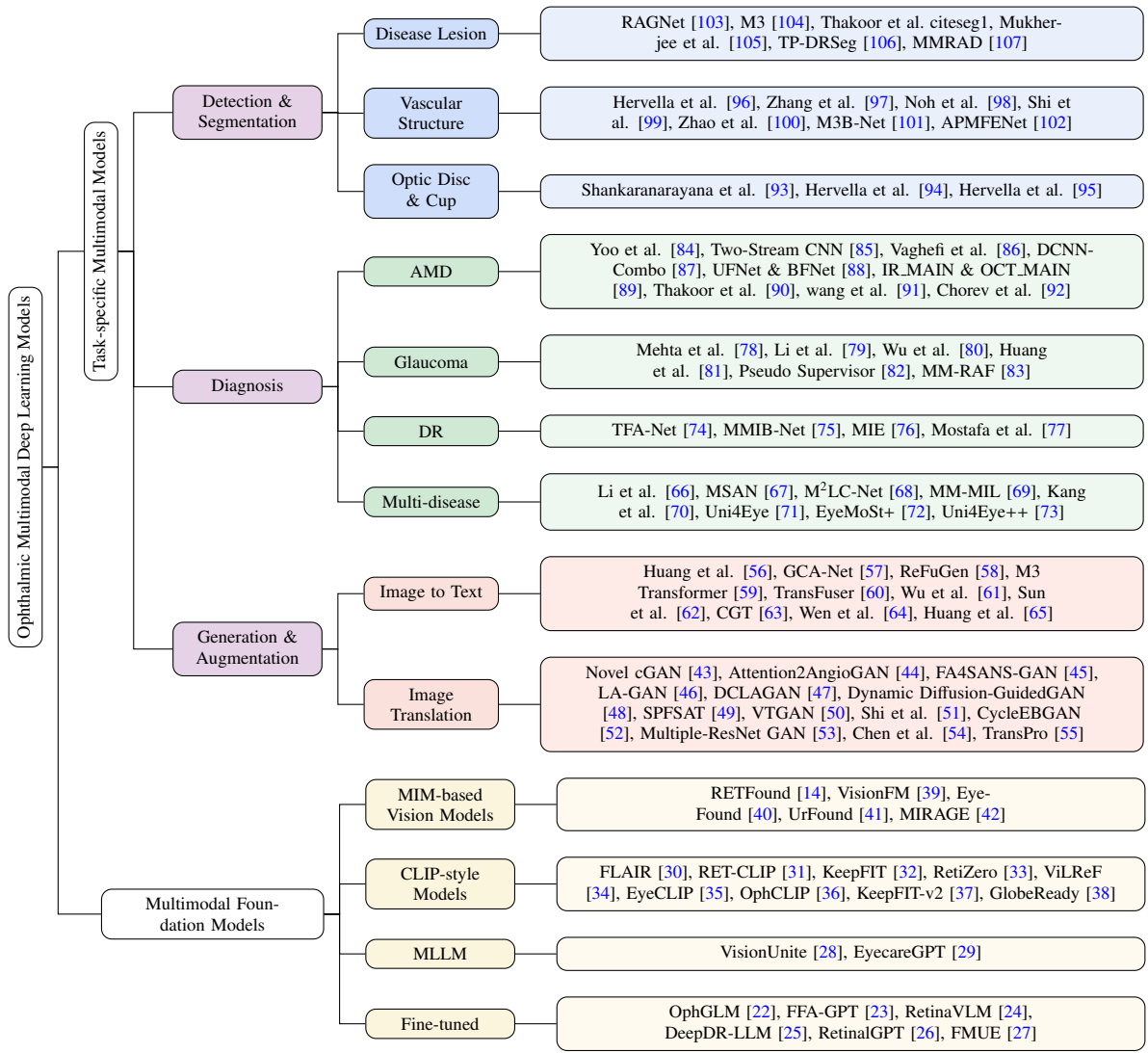
\begin{figure*}[t]
    \centering

\begin{forest}
    for tree={
        grow=east,
        draw,
        rounded corners,
        node options={align=center, font=\scriptsize},
        edge path={
            \noexpand\path[\forestoption{edge}]
            (!u.south) -- ++(6pt,0) |- (.north);
        },
        l sep=16pt,
        s sep=10pt,
        anchor = west,
    }
    [Ophthalmic Multimodal Deep Learning Models,
        rotate=90,
        [Multimodal Foundation Models,
            rotate=0,
            text width=2.8cm,
            edge path={
                \noexpand\path[\forestoption{edge}]
                (!u.south) -- ++(6pt,0) |- (.west);
            },
            for children={
                rotate=0,
                text width=1.8cm,
                edge path={
                    \noexpand\path[\forestoption{edge}]
                    (!u.east) -- ++(6pt,0) |- (.west);
                },
                fill=colorFM,
            }
            [Fine-tuned,
                calign=child edge,
                for children={
                    edge path={
                        \noexpand\path[\forestoption{edge}]
                        (!u.east) -- (.west);
                    },
                }
                [{OphGLM \cite{ophglm}, FFA-GPT \cite{ffagpt}, RetinaVLM \cite{retinalvlm}, DeepDR-LLM \cite{deepdr}, RetinalGPT \cite{retinalgpt}, FMUE \cite{fmue}},
                text width=7.8cm,
                fill=colorFM2,]
            ]
            [MLLM,
                calign=child edge,
                for children={
                    edge path={
                        \noexpand\path[\forestoption{edge}]
                        (!u.east) -- (.west);
                    },
                }
                [{VisionUnite \cite{visionunite}, EyecareGPT \cite{eyecaregpt}},
                text width=7.8cm,
                fill=colorFM2,]
            ]
            [CLIP-style Models,
                calign=child edge,
                for children={
                    edge path={
                        \noexpand\path[\forestoption{edge}]
                        (!u.east) -- (.west);
                    },
                }
                [{FLAIR \cite{flair}, RET-CLIP \cite{retclip}, KeepFIT \cite{mmretinal},  RetiZero \cite{retizero}, ViLReF \cite{vilref}, EyeCLIP \cite{eyeclip}, OphCLIP \cite{ophclip}, KeepFIT-v2 \cite{mmretinalv2}, GlobeReady \cite{globeready}},
                text width=7.8cm,
                fill=colorFM2,]
            ]
            [MIM-based Vision Models,
                calign=child edge,
                for children={
                    edge path={
                        \noexpand\path[\forestoption{edge}]
                        (!u.east) -- (.west);
                    },
                }
                [{RETFound \cite{intro14}, VisionFM \cite{visionfm}, EyeFound \cite{eyefound}, UrFound \cite{urfound}, MIRAGE \cite{mirage}},
                text width=7.8cm,
                fill=colorFM2,]
            ]
        ]
        [Task-specific Multimodal Models,
            rotate=90,
            for children={
                rotate=0,
                text width=1.8cm,
                edge path={
                    \noexpand\path[\forestoption{edge}]
                    (!u.south) -- ++(6pt,0) |- (.west);
                },
                fill=colorTit1,
            }
            [Generation \& Augmentation,
                for children={
                    edge path={
                        \noexpand\path[\forestoption{edge}]
                        (!u.east) -- ++(6pt,0) |- (.west);
                    },
                    text width=1.6cm,
                    fill=colorGer1,
                }
                [Image Translation,
                    calign=child edge,
                    for children={edge path={
                        \noexpand\path[\forestoption{edge}]
                        (!u.east) -- (.west);
                    },}
                    [{Novel cGAN \cite{ger12}, Attention2AngioGAN \cite{ger13}, FA4SANS-GAN \cite{ger14}, LA-GAN \cite{ger15}, DCLAGAN \cite{ger16}, Dynamic Diffusion-GuidedGAN \cite{ger17}, SPFSAT \cite{ger18}, VTGAN \cite{ger19}, Shi et al. \cite{ger20}, CycleEBGAN \cite{ger21}, Multiple-ResNet GAN \cite{ger22}, Chen et al. \cite{ger23}, TransPro \cite{ger24}},
                    text width=8cm,
                    fill=colorGer2,]
                ]
                [Image to Text,
                    calign=child edge,
                    for children={edge path={
                        \noexpand\path[\forestoption{edge}]
                        (!u.east) -- (.west);
                    },}
                    [{Huang et al. \cite{ger1}, GCA-Net \cite{ger2}, ReFuGen \cite{ger3}, M3 Transformer \cite{ger4}, TransFuser \cite{ger5}, Wu et al. \cite{ger6}, Sun et al. \cite{ger7}, CGT \cite{ger8}, Wen et al. \cite{ger10}, Huang et al. \cite{ger11}},
                    text width=8cm,
                    fill=colorGer2,]
                ]
            ]
            [Diagnosis,
                for children={
                    edge path={
                        \noexpand\path[\forestoption{edge}]
                        (!u.east) -- ++(6pt,0) |- (.west);
                    },
                    text width=1.6cm,
                    fill=colorDig1,
                }
                [Multi-disease,
                    calign=child edge,
                    for children={edge path={
                        \noexpand\path[\forestoption{edge}]
                        (!u.east) -- (.west);
                    },}
                    [{Li et al. \cite{Multi2020_LiXiaomengi}, MSAN \cite{Multi2021_MSAN}, M\textsuperscript{2}LC-Net \cite{Multi2021_M2LC_Net}, MM-MIL \cite{Multi2021_MM_MIL}, Kang et al. \cite{Multi2021_Kang}, Uni4Eye \cite{Multi2022_Uni4Eye}, EyeMoSt+ \cite{Multi2024_EyeMost_plus}, Uni4Eye++ \cite{Multi2024_Uni4eye_plusplus}},
                    text width=8cm,
                    fill=colorDig2,]
                ]
                [DR,
                    calign=child edge,
                    for children={edge path={
                        \noexpand\path[\forestoption{edge}]
                        (!u.east) -- (.west);
                    },}
                    [{TFA-Net \cite{DR2020_TFA_Net}, MMIB-Net \cite{DR2021_MMIB_Net}, MIE \cite{DR2022_MIE}, Mostafa et al. \cite{DR2023_Mostafa}},
                    text width=8cm,
                    fill=colorDig2,]
                ]
                [Glaucoma,
                    calign=child edge,
                    for children={edge path={
                        \noexpand\path[\forestoption{edge}]
                        (!u.east) -- (.west);
                    },}
                    [{Mehta et al. \cite{Glau2021_DenseNet}, Li et al. \cite{Glau2022_}, Wu et al. \cite{Glau2023_GAMMA}, Huang et al. \cite{Glau2023_GRAPE}, Pseudo Supervisor \cite{Glau2023_Pseudo_Supervisor}, MM-RAF \cite{Glau2023_MM_RAF}},
                    text width=8cm,
                    fill=colorDig2,]
                ]
                [AMD,
                    calign=child edge,
                    for children={edge path={
                        \noexpand\path[\forestoption{edge}]
                        (!u.east) -- (.west);
                    },}
                    [{Yoo et al. \cite{AMD2018_RF}, Two-Stream CNN \cite{AMD2019_MM_CNN}, Vaghefi et al. \cite{AMD2020_Inception_ResNet_V2_Based_MM_CNN}, DCNN-Combo \cite{AMD2021_DCNN_Combo}, UFNet \& BFNet \cite{AMD2022_XFNet}, IR\_MAIN \& OCT\_MAIN \cite{AMD2022_X_MAIN}, Thakoor et al. \cite{AMD2022_Hybrid_CNN}, wang et al. \cite{AMD2022_Wang}, Chorev et al. \cite{AMD2023_25DResNet50}},
                    text width=8cm,
                    fill=colorDig2,]
                ]
            ]
            [Detection \& Segmentation,
                for children={
                    edge path={
                        \noexpand\path[\forestoption{edge}]
                        (!u.east) -- ++(6pt,0) |- (.west);
                    },
                    text width=1.6cm,
                    fill=colorSeg1,
                }
                [Optic Disc \& Cup,
                    calign=child edge,
                    for children={edge path={
                        \noexpand\path[\forestoption{edge}]
                        (!u.east) -- (.west);
                    },}
                    [{Shankaranarayana et al. \cite{seg14}, Hervella et al. \cite{seg15}, Hervella et al. \cite{seg16}},
                    text width=8cm,
                    fill=colorSeg2,]
                ]
                [Vascular Structure,
                    calign=child edge,
                    for children={edge path={
                        \noexpand\path[\forestoption{edge}]
                        (!u.east) -- (.west);
                    },}
                    [{Hervella et al. \cite{seg7}, Zhang et al. \cite{seg8}, Noh et al. \cite{seg9}, Shi et al. \cite{seg10}, Zhao et al. \cite{seg11}, M3B-Net \cite{seg12}, APMFENet \cite{seg13}},
                    text width=8cm,
                    fill=colorSeg2,]
                ]
                [Disease Lesion,
                    calign=child edge,
                    for children={edge path={
                        \noexpand\path[\forestoption{edge}]
                        (!u.east) -- (.west);
                    },}
                    [{RAGNet \cite{seg6}, M3 \cite{seg3}, Thakoor et al.\ cite{seg1}, Mukherjee et al. \cite{seg2}, TP-DRSeg \cite{seg4}, MMRAD \cite{seg5}},
                    text width=8cm,
                    fill=colorSeg2,]
                ]
            ]
        ]
    ]
\end{forest}

\caption{Taxonomy of multimodal models in ophthalmology.}

\label{fig2}

\end{figure*}

\subsubsection{Multimodal Foundation Models}

Multimodal foundation models are large-scale AI systems pretrained on vast amounts of unlabeled and heterogeneous data from multiple modalities. This extensive pretraining enables them to generalize effectively and adapt to a wide range of downstream ophthalmic tasks with minimal task-specific fine-tuning. Based on their data fusion strategies and application scenarios, the current applications of these models in ophthalmology can be broadly categorized into three types: 1) \textbf{multimodal vision models}, which integrate diverse imaging modalities such as color fundus photography (CFP), optical coherence tomography (OCT), and fluorescein angiography (FA) to enhance diagnostic accuracy and enable more comprehensive disease assessments; 2) \textbf{CLIP-based models}, which utilize contrastive learning to align ophthalmic images with corresponding clinical text reports, thereby improving cross-modal retrieval performance and model interpretability; and 3) \textbf{multimodal large language models (LLMs)}, which combine structured and unstructured clinical data with medical images to support complex tasks such as diagnostic report generation, disease progression prediction, and natural language-based clinical decision support.

\begin{table*}[t]
\centering
\fontsize{7}{9}\selectfont
\caption{Comparison of ophthalmic imaging modalities.}
\label{modalities}
\renewcommand{\arraystretch}{1.3}
\begin{tabularx}{\textwidth}{%
  >{\centering\arraybackslash}p{0.8cm}  
  >{\centering\arraybackslash}p{3.0cm}  
  >{\centering\arraybackslash}p{2.3cm}  
  >{\centering\arraybackslash}p{1.3cm}  
  >{\arraybackslash}p{6.5cm}            
}
\toprule
\textbf{Abbr.} & \textbf{Full Name} & \textbf{Imaged Region} & \textbf{Clinical Use} & \textbf{Typical Application} \\
\midrule
CFP & Color Fundus Photography & Retinal surface & Yes & Screening and documentation of retinal diseases \cite{rajalakshmi2021review}\\

OCT & Optical Coherence Tomography & Retina, macula, optic nerve & Yes & High-resolution layered retinal imaging, diagnosis of macular edema and glaucoma \cite{bouma2022optical}\\

OCTA & OCT Angiography & Retinal and choroidal vasculature & Yes & Non-invasive blood flow visualization and neovascularization assessment \cite{spaide2018optical}\\

FFA & Fundus Fluorescein Angiography & Retinal vessels & Limited & Dynamic blood flow observation, neovascularization, leakage diagnosis \cite{klufas2015influence}\\

Slit-Lamp & Slit-Lamp Biomicroscopy & Cornea, anterior chamber, lens & Yes & Anterior segment evaluation for cataract, keratitis, trauma \cite{YANG2025103533}\\

ICGA & Indocyanine Green Angiography & Choroidal vessels & Limited  & Deep choroidal vessel imaging (e.g., PCV, choroidal neovascularization) \cite{chen2024translating}\\

OUS & Ocular Ultrasound & Posterior segment, vitreous & Yes & Evaluation through opaque media, detects retinal detachment, vitreous hemorrhage \cite{silverman2023principles}\\

FAF & Fundus Autofluorescence & RPE metabolic status & Yes & Detects RPE-related diseases (e.g., Stargardt, geographic atrophy) \cite{schmitz2021fundus}\\

IR & Infrared Reflectance Imaging & Retina, choroid & Yes  & Identifies dark lesions and supports OCT/OCTA interpretation \cite{saleh2022role}\\

UWF & Ultra-Widefield Imaging & Peripheral retina & Yes & Comprehensive DR, retinal tear, vein occlusion evaluation \cite{patel2020ultra}\\

MRI & Magnetic Resonance Imaging & Orbit, optic pathway & Limited & Orbital tumors, optic nerve lesions, central nervous system involvement \cite{murthy2016clinical}\\

UBM & Ultrasound Biomicroscopy & Cornea, anterior chamber, ciliary body & Yes & Glaucoma mechanism, angle and anterior segment imaging \cite{foster2000advances}\\

LSO & Laser Scanning Ophthalmoscopy & Retinal surface & Yes & Often combined with OCT for composite imaging \cite{akyol2021adaptive}\\

B-Scan & B-Scan Ultrasonography & Vitreous, posterior segment & Yes & Retinal detachment and tumor detection when media are opaque \cite{kadakia2023ultrasound}\\

Specular & Specular Microscopy & Corneal endothelium & Yes & Endothelial cell count, essential before and after corneal transplantation \cite{mohamed2016corneal}\\

External & External Eye Photography & Eyelid, conjunctiva, corneal surface & Yes & Documentation of anterior segment disease and perioperative changes \cite{babenko2022detection}\\

Topo & Corneal Topography & Corneal curvature & Yes & Keratoconus screening and pre-refractive surgery planning \cite{kanclerz2021current}\\

RetCam & Retinal Camera for Infants & Infant retina & Yes & ROP (Retinopathy of Prematurity) screening \cite{goyal2018outcome}\\

FS & Fluorescein Staining Imaging & Cornea, conjunctiva & Yes & Detect corneal epithelial defects, dry eye, ulcers, and foreign bodies \cite{pellegrini2019assessment}\\

CT & Computed Tomography & Orbit, bone, sinus region & Limited & Orbital fractures, calcified lesions, intraocular foreign bodies \cite{chalam2016optical}\\
\bottomrule
\end{tabularx}
\end{table*}

\subsection{Scope and Modalities of This Review}
This review focuses on ophthalmic multimodal deep learning research published between 2018 and 2025. In the Task-Specific Multimodal Models section, we exclusively include peer-reviewed and accepted papers, ensuring the reliability and maturity of the reported methods. In contrast, the Multimodal Foundation Models section also incorporates preprint articles (e.g., arXiv), reflecting the rapid development and emerging nature of foundation model research in ophthalmology. This inclusion aims to capture the cutting-edge progress in this evolving area, despite the lack of formal peer review in some cases. A comprehensive tree diagram summarizing all the methods reviewed in this paper is provided in Figure \ref{fig2}, offering a structured view of the included works and their categorizations.

In ophthalmic research, data sources span both visual and non-visual modalities. Among non-image modalities, textual data plays a vital role and primarily includes clinical reports and visual field (VF) measurements. Clinical reports often contain diagnostic impressions, treatment history, and narrative descriptions, while VF data quantitatively reflect the functional status of a patient's visual system, especially in conditions like glaucoma. Ophthalmic imaging, on the other hand, encompasses a broad spectrum of modalities tailored for different anatomical structures and pathological assessments. This review considers 17 well-defined imaging modalities, covering fundus photography, OCT, angiography, and ultra-widefield techniques, among others. A detailed listing of all involved modalities is provided in Table \ref{modalities}.

\subsection{Related Work}
In recent years, multiple reviews have examined deep learning applications in ophthalmology, with most focusing on single-modal data, especially fundus images. One representative work covers tasks such as lesion detection, vessel and optic disc/cup segmentation, disease classification, and image synthesis. It also highlights common challenges like limited annotations, class imbalance, and poor generalizability, proposing weak supervision and domain adaptation as potential solutions \cite{rel1}.

Another study reviewed CNN-based methods for segmentation and classification, emphasizing architectures like U-Net, attention mechanisms, and adversarial learning. It also discussed preprocessing strategies and model deployment issues \cite{rel2}. A focused survey on retinal vessel segmentation categorized 110 studies using U-Net, FCNs, and GANs, pointing out difficulties such as noisy images, vessel heterogeneity, and lack of standardization \cite{rel3}.

Beyond segmentation, several reviews have investigated the deep learning methods for diagnosing major eye diseases such as diabetic retinopathy, glaucoma, age-related macular degeneration, and retinopathy of prematurity using fundus and OCT images. These studies also explored challenges, including data imbalance and model interpretability, while applying convolutional neural networks, transfer learning, and generative adversarial network-based synthesis \cite{rel4}.

Further, some surveys started to include multimodal elements, such as combining fundus and OCT for disease classification, or integrating demographic and clinical variables for personalized prediction \cite{rel5}. Most recently, new work has emerged on vision-language models (VLMs) and foundation models that jointly process images and clinical texts, enabling automated reporting and disease interpretation. However, these efforts remain scattered and lack a focused review \cite{rel6}.

In summary, while existing literature has provided valuable insights into unimodal deep learning, there is still no comprehensive survey dedicated to multimodal deep learning in ophthalmology. To address this gap, this review synthesizes recent progress (2018–2025), covering both task-specific multimodal models and foundation models, aiming to outline key methods, challenges, and future directions in this rapidly evolving area. In total, this survey examines 64 task-specific multimodal studies and 13 works on foundation models, providing a broad yet detailed overview of the field’s development.

\begin{figure*}[t]
\centering
\includegraphics[width=1\textwidth]{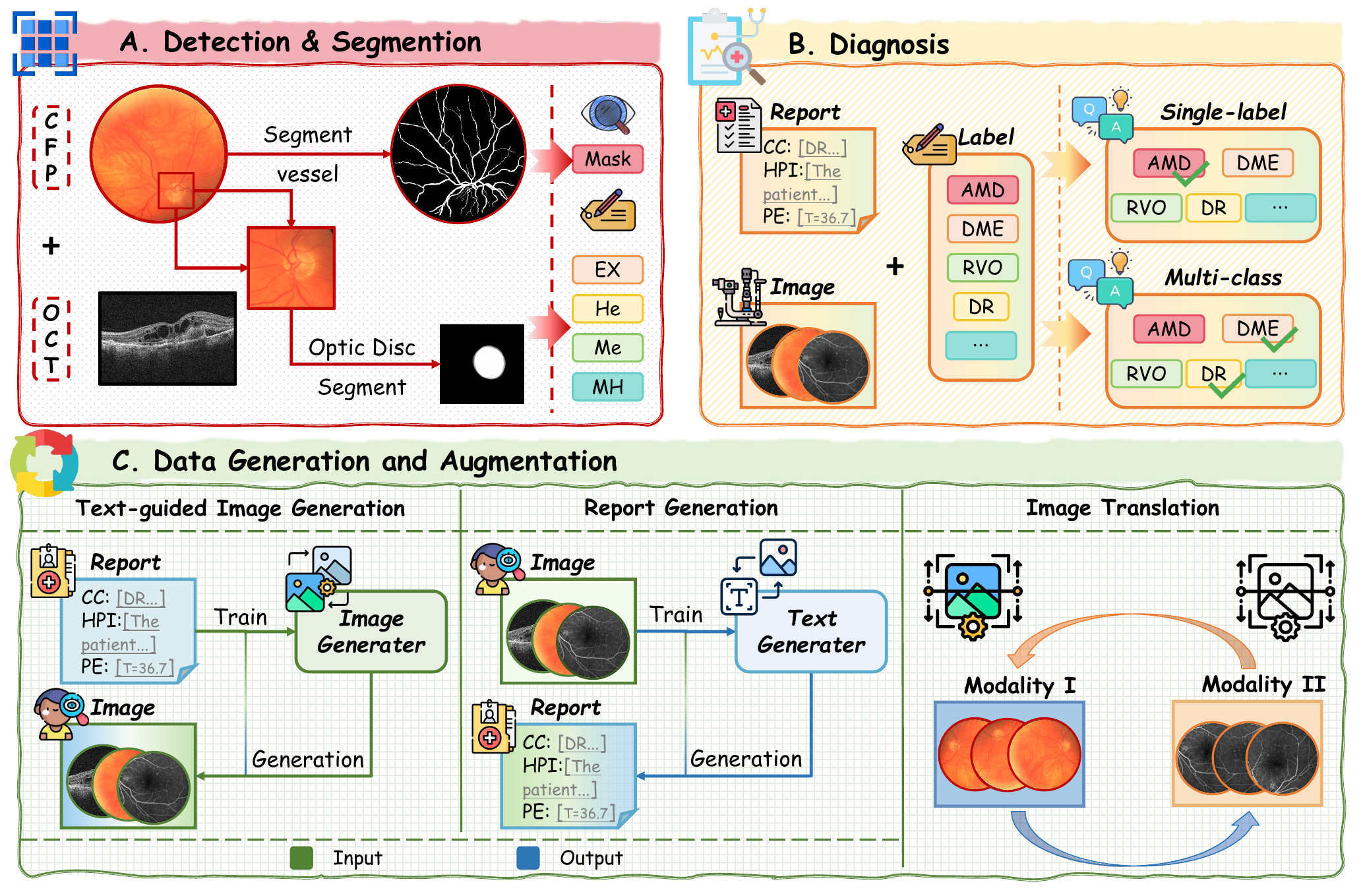}
\caption{Core directions of ophthalmic multimodal modeling: detection, diagnosis, and generation.}
\label{fig3}
\end{figure*}

\section{Task-specific Multimodal Models in Ophthalmology}
\label{sec3}

In ophthalmic artificial intelligence research, multimodal deep learning models excluding foundation models are generally developed with a focus on specific clinical tasks. Due to the complexity of ophthalmic diseases and the diversity of data modalities such as color fundus photography, optical coherence tomography (OCT), and visual field tests, model architectures and optimization objectives are frequently tailored to address particular clinical challenges. This task-specific strategy aims to deliver precise and efficient clinical outcomes. Accordingly, the existing literature can be divided into three main groups. The first concerns disease diagnosis, which emphasizes pathological screening and grading through classification models. The second addresses lesion detection and segmentation, facilitating quantitative analysis through localization and contour extraction. The third involves data generation, employing generative models to alleviate data scarcity and improve image quality. As illustrated in Figure \ref{fig3}, these groups collectively encompass the essential dimensions of ophthalmic multimodal modeling tasks. The following sections provide a detailed discussion of the technical features and recent progress within each of these areas.

\subsection{Multimodal Ophthalmic Image Detection and Segmentation}

Multimodal ophthalmic imaging, which includes optical coherence tomography (OCT), color fundus photography (CFP), fluorescein and indocyanine-green angiography (FA and ICGA), as well as related modalities, provides complementary structural and functional information. These combined cues significantly improve the detection and segmentation of retinal lesions, vascular networks, and the optic-disc and cup complex. Recent advances in deep learning that integrate cross-modal feature fusion through methods such as attention mechanisms and vision–language alignment, along with semi-supervised strategies like generative pretraining and label transfer, and multi-task joint learning, have helped to overcome challenges related to data scarcity and modality heterogeneity while enhancing model generalizability. As summarized in Table \ref{Lesion-Detection}, this section offers a systematic survey of the latest progress in multimodal models for ophthalmic image analysis and emphasizes how the complementary use of modalities, data-efficient training approaches, and robust transferability are driving these techniques closer to clinical application.

\begin{table*}[t]
\centering
\fontsize{7}{9}\selectfont
\caption{Summary of multimodal approaches for ophthalmic lesion detection and segmentation.}
\label{Lesion-Detection}
\begin{tabularx}{\textwidth}{%
  >{\RaggedRight\arraybackslash}p{0.4cm}  
  >{\RaggedRight\arraybackslash}p{2.2cm}  
  >{\RaggedRight\arraybackslash}p{1.0cm}  
  >{\RaggedRight\arraybackslash}p{4.4cm}  
  >{\RaggedRight\arraybackslash}p{1.8cm}  
  >{\RaggedRight\arraybackslash}p{3.8cm}  
}
\toprule
\textbf{Year} & \textbf{Study} & \textbf{Approach} & \textbf{Application} & \textbf{Modality} & \textbf{Dataset} \\
\toprule
\multicolumn{6}{@{}p{\textwidth}@{}}{\quad Lesion Detection and Segmentation}\\
\midrule
2020 & Hassan et al. \cite{seg6} & CNN & Multi-lesion segmentation & CFP, OCT & Rabbani-I/II, Duke-I/II/III, BIOMISA, Kermany et al. 2018 \\
2021 & Chen et al. \cite{seg3} & CNN & RPD detection, AMD classification & CFP, FAF & AREDS2, Rotterdam Study \\
2022 & Thakoor et al. \cite{seg1} & CNN & Biomarker detection, AMD classification & OCTA, OCT, 2D/3D B-scans & Private \\
2022 & Mukherjee et al. \cite{seg2} & UNet & Drusen segmentation & CFP, OCT & Private \\
2024 & Li et al. \cite{seg4} & VLM + SAM & DR lesion segmentation & CFP, Description & IDRiD, DDR \\
2024 & Li et al. \cite{seg5} & SAM & RAO lesions detection and localization & CFP, OCT & MMR \\

\midrule
\multicolumn{6}{@{}p{\textwidth}@{}}{\quad Vascular Structure Segmentation}\\
\midrule
2019 & Hervella et al. \cite{seg7} & UNet & Vessel segmentation & CFP, FFA & DRIVE, STARE, Isfahan MISP \\
2019 & Zhang et al. \cite{seg8} & CNN & Vessel segmentation & CFP, FFA & Private \\
2020 & Noh et al. \cite{seg9} & CNN & Vessel segmentation & CFP, FFA & SNUBH Fundus-FA (Private) \\
2024 & Shi et al. \cite{seg10} & GAN & Artery and vein segmentation & CFP, FFA & RITE, HRF, LES-AV, AV-WIDE, PortableAV (Private), DRSplusAV (Private) \\
2024 & Zhao et al. \cite{seg11} & CNN& Vessel segmentation & OCTA (2D \& 3D) & OCTA-500, ROSE \\
2025 & Xie et al. \cite{seg12} & CNN & Vessel segmentation & UWF, FFA & UWF-SEG (Private), PRIME-FP20 \\
2025 & Quan et al. \cite{seg13} & UNet & Vessel and FAZ segmentation & OCT, OCTA & OCTA-500 \\

\midrule
\multicolumn{6}{@{}p{\textwidth}@{}}{\quad Optic Disc and Cup Segmentation}\\
\midrule
2019 & Shankaranarayana et al. \cite{seg14} & FCN & Optic Disc \& Cup Segmentation & CFP, FFA & INSPIRE-stereo, ORIGA, RIMONE, DRISHTI-GS \\
2020 & Hervella et al. \cite{seg15} & UNet & Optic Disc \& Cup Segmentation & CFP, FFA & Isfahan MISP, DRISHTI-GS, REFUGE \\
2022 & Hervella et al. \cite{seg16} & UNet & Optic Disc \& Cup Segmentation, Glaucoma Classification & CFP, FFA & REFUGE, DRISHTI-GS \\

\bottomrule
\end{tabularx}
\end{table*}

\subsubsection{Lesion Detection and Segmentation}
Recent work demonstrates that carefully designed fusion frameworks can exploit the complementary strengths of different imaging modalities to locate and delineate retinal lesions more reliably than single-source models. Thakoor et al. \cite{seg1} introduced a 2D–3D hybrid CNN that concatenates features from OCT angiography (OCTA), structural OCT slices and B-scan flow maps through long skip connections. This model achieved 70.8\% accuracy and an AUC of 98.50\% in the three-class diagnosis of age-related macular degeneration (AMD), underscoring the value of volumetric cues. Mukherjee et al. \cite{seg2} adopted a Siamese architecture to fuse stereo colour fundus photographs (CFP) with an OCT-derived RPE-elevation map. Their application of the STAPLE algorithm improved the drusen-segmentation Dice score to 72.00\%, indicating that low-cost CFP can be leveraged to infer high-value OCT parameters. To integrate functional signals, Chen et al. \cite{seg3} proposed M3, which couples self-attention with cross-modal attention to merge CFP and fundus autofluorescence (FAF). Even when using input from a single modality, it achieved an external test F1 score of 78.74\% for RPD segmentation, outperforming retinal specialists. Moving beyond images alone, TP-DRSeg \cite{seg4} combined vision and language: CLIP text prompts such as “hard exudate” injected explicit priors via a prior-alignment injector to boost SAM-based segmentation (mDice 49.72\%), alleviating reliance on hand-crafted prompts.

Semi-supervised paradigms have also proved useful. MMRAD \cite{seg5} built on SAM to detect retinal artery occlusion (RAO) across CFP and OCT by sharing a decoder and task-specific tokens. Its low-level prompt-tuning strategy required only healthy images with simulated anomalies and still reached an AUC of 76.64\%. Finally, RAGNet \cite{seg6} unified lesion segmentation and pathology classification within a mixed convolutional backbone, sharing a ResNet-50 encoder; on heterogeneous OCT and CFP datasets, it attained a mean Dice of 82.20\% and cross-dataset IoU of 71.20\%, illustrating that joint learning can lift single-task performance.

\subsubsection{Vascular Structure Segmentation}

Multimodal vascular analysis has made significant progress through the integration of wide-field, depth-resolved, and perfusion-sensitive signals. Hervella et al. \cite{seg7} is the first to explore the potential of fluorescein angiography (FFA) for self-supervised learning by auto-generating vessel labels. They trained a network that achieved an AUC-ROC of 97.40\% on the DRIVE dataset, which is comparable to fully supervised methods while greatly reducing annotation costs. Building on this work, Zhang et al. \cite{seg8} jointly trained a vessel segmentation and deformation registration network that incorporated a style-transfer module to improve modality alignment. Their approach achieved a Dice score of 65.46\% without relying on ground-truth deformation fields. More recently, M3B-Net \cite{seg12}, a multi-branch U-Net architecture, was introduced to process ultra-wide-field (UWF) and FFA images and included CycleGAN-based style-transfer enhancement. This design featured a selective-fusion module and a local-perception fusion module, leading to a Dice score of 83.02\% on the PRIME-FP20 dataset.

Three-dimensional modalities have brought additional performance gains to retinal vessel analysis. Zhao et al. \cite{seg11} developed a local-and-global mutual learning framework that integrates 3D OCT/OCTA volumes with their 2D projections, achieving a Dice score of 93.43\% on the OCTA-500 dataset. Building on this direction, APMFENet \cite{seg13} introduced kernel-adaptive asymmetric convolutions and a multi-scale context fusion module (MCFM) to compress 3D volumes efficiently. This enabled simultaneous segmentation of retinal vessels and the foveal avascular zone (FAZ), reaching Dice scores of 89.96\% and 97.17\%, respectively.

When labeled data is limited, weakly supervised or unsupervised approaches can still achieve strong performance. Noh et al. \cite{seg9} proposed a hierarchical registration pipeline to fuse color fundus photography (CFP) with fluorescein angiography (FFA), significantly enhancing vessel detection. Their method achieved a sensitivity of 99.40\% and even outperformed specialists in identifying fine, thread-like vessels. More recently, Shi et al. \cite{seg10} leveraged arterial–venous phase shifts in FFA to generate soft labels. After a GAN-based pre-training phase, the model was fine-tuned using just a single labeled sample, achieving an arterial Dice score of 77.30\% with specificity exceeding 98.00\%.
\subsubsection{Optic Disc and Cup Segmentation}
Depth cues and cross-modal pre-training have revitalised optic-disc and cup (OD/OC) analysis. Shankaranarayana et al. \cite{seg14} generated monocular pseudo-depth maps from RGB fundus images via self-supervised pre-training and merged them with colour features in a dilated-residual-inception (DRI) block.Experimental results on the ORIGA dataset demonstrated that incorporating depth information improved the cup segmentation Dice coefficient to 87.60\% and reduced the cup-to-disc ratio (CDR) error to 6.70\%. Hervella et al. \cite{seg15} adopted multimodal self-supervision between CFP and fluorescein angiography (FA) to improve OD/OC delineation when labels are sparse, lifting the cup Jaccard index on DRISHTI-GS from 75.36\% to 82.29\%. Extending the idea, Hervella et al. \cite{seg16} showed that multimodal pre-training (CFP + FA) followed by single-modal fine-tuning in a multi-task framework (segmentation + classification) produced a cup Dice of 91.03\% and glaucoma-classification AUC of 94.74\% on DRISHTI-GS, revealing the promise of a “pre-train multimodal, fine-tune unimodal” paradigm.

\begin{table*}[t]
\centering
\fontsize{7}{9}\selectfont  
\caption{Summary of multimodal approaches in ophthalmic disease diagnosis.}
\label{disease-diagnosis}
\begin{tabularx}{\textwidth}{llllll}
\toprule
\textbf{Year} & \textbf{Study} & \textbf{Approach} & \textbf{Application} & \textbf{Modality} & \textbf{Dataset} \\
\toprule
\multicolumn{6}{@{}p{\textwidth}@{}}{\quad AMD}\\
\midrule
2018 &Yoo et al. \cite{AMD2018_RF} &CNN &AMD classification &CFP, OCT &Project Macula \\
2019 &Wang et al. \cite{AMD2019_MM_CNN} &CNN &AMD classification &CFP, OCT &Private \\
2020 &Vaghefi et al. \cite{AMD2020_Inception_ResNet_V2_Based_MM_CNN} &CNN &AMD classification &CFP, OCT, OCT-A &Private \\
2021 &Xu et al. \cite{AMD2021_DCNN_Combo} &CNN &AMD and PCV classification &CFP, OCT &Private \\
2022 &Jin et al. \cite{AMD2022_XFNet} &CNN &CNV indentification &OCT, OCT-A &Private \\
2022 &Chen et al. \cite{AMD2022_X_MAIN} &CNN &AMD classification &OCT, IR &Private \\
2022 &Thakoor et al. \cite{AMD2022_Hybrid_CNN} &CNN &AMD classification &OCT, OCT-A, B-Scan, OCT &Private \\
2022 &Wang et al. \cite{AMD2022_Wang} &CNN &AMD classification &CFP, OCT &Private \\
2023 &Chorev et al. \cite{AMD2023_25DResNet50} &CNN &AMD classification &clinical data, OCT &Private \\
\midrule
\multicolumn{6}{@{}p{\textwidth}@{}}{\quad Glaucoma}\\
\midrule
2021 &Mehta et al. \cite{Glau2021_DenseNet} &CNN &Glaucoma detection &Demographic information &Upon Request \\
     &                    &    &                   &clinical data, CFP, OCT & \\
2022 &Li et al. \cite{Glau2022_} &CNN &Glaucoma and PDR classification &CFP, OCT, OCT-A, LSO &GAMMA \\
2022 &Wu et al. \cite{Glau2023_GAMMA} &- &Glaucoma classification &CFP, OCT &GAMMA \\
2022 &Huang et al. \cite{Glau2023_GRAPE} &CNN &Glaucoma classification &CFP, OCT, VF, clinical data &GRAPE \\
2022 &Luo et al. \cite{Glau2023_Pseudo_Supervisor} &pseudo supervisor &Glaucoma detection and &OCT, VF, 
Demographic information &Harvard \\
     &              &                  &progression forecasting  & & \\
2022 &Zhou et al. \cite{Glau2023_MM_RAF} &Transformer &Glaucoma recognition &CFP, OCT & GAMMA \\

\midrule
\multicolumn{6}{@{}p{\textwidth}@{}}{\quad DR}\\
\midrule
2020 &Hua et al. \cite{DR2020_TFA_Net} &CNN &DR Grading &CFP, OCTA &MESSIDOR \\
2021 &Song et al. \cite{DR2021_MMIB_Net} &CNN &DR Detection &BR, GR, IR, Combined-pseudocolor &Private \\
2022 &Hervella et al. \cite{DR2022_MIE} &CNN &DR Grading &CFP, FFA &Public \\
2023 &Mostafa et al. \cite{DR2023_Mostafa} &CNN &DR Grading &UWF-CFP, OCTA &Private \\

\midrule
\multicolumn{6}{@{}p{\textwidth}@{}}{\quad Multi-disease}\\
\midrule
2020 &Li et al. \cite{Multi2020_LiXiaomengi} &CNN &Retinal disease classification &CFP, FFA &Public \\
2021 &He et al. \cite{Multi2021_MSAN} &CNN, Transformer &Retinal disease classification &CFP, OCT &Private \\
2021 &Ou et al. \cite{Multi2021_M2LC_Net} &CNN, Transformer &Retinal disease classification &CFP, OCT &Private \\
2021 &Li et al. \cite{Multi2021_MM_MIL} &CNN &Retinal disease recognition &CFP, OCT &Private \\
2021 &Kang et al. \cite{Multi2021_Kang} &CNN &Retinal vascular disease Testing &CFP, OCT, FA/ICGA &Private \\
2022 &Cai et al. \cite{Multi2022_Uni4Eye} &Transformer &Retinal disease classification &CFP, OCT, OCT-A, FFA &Public \\
2024 &Zou et al. \cite{Multi2024_EyeMost_plus} &Transformer &Retinal disease classification &CFP, OCT &Public \\
2024 &Cai et al. \cite{Multi2024_Uni4eye_plusplus} &Transformer &Retinal disease classification &CFP, OCT, OCT-A, FFA &Public \\

\bottomrule
\end{tabularx}
\end{table*}

\subsection{Multimodal Diagnostic Models for Ophthalmic Diseases}
\subsubsection{Age-related Macular Degeneration Diagnosis}

Age-related macular degeneration (AMD) is a degenerative disease that affects the central region of the retina known as the macula and is mainly classified into two types: dry and wet \cite{AMD}. Dry AMD presents with drusen, whereas wet AMD involves leakage caused by choroidal neovascularization (CNV). Early stages of AMD often have no symptoms and traditional imaging techniques have limited ability to reveal disease features, which makes clinical diagnosis challenging due to low sensitivity and inaccurate localization \cite{AMD2020_Inception_ResNet_V2_Based_MM_CNN, AMD2021_DCNN_Combo, AMD2022_X_MAIN, AMD2022_Hybrid_CNN, AMD2022_Wang, AMD2023_25DResNet50}. Furthermore, evaluating CNV activity is essential for guiding treatment decisions, but non-invasive and highly precise assessment methods are currently lacking.

In recent years, several studies have attempted to use multimodal deep learning models to improve the diagnostic accuracy of AMD. The works of the related ophthalmological diagnostics are listed in Table \ref{disease-diagnosis}. For example, Yoo et al. \cite{AMD2018_RF} proposed a multimodal model based on a pre-trained VGG-19 network combined with transfer learning using a random forest, integrating CFP and OCT image information to improve the classification performance of AMD. Wang et al. \cite{AMD2019_MM_CNN} further proposed a method combining a two-stream convolutional neural network (Two-stream CNN) with a loose pairing training strategy to address the issue of insufficient multimodal training data, and used multi-modal class activation maps (CAM) to visually demonstrate the contribution of each modality to the final prediction results. Jin et al. \cite{AMD2022_XFNet}developed a feature-level multi-modal model based on the fusion of OCT and OCTA images for identifying CNV activity, which demonstrated high diagnostic accuracy in both internal and external validation datasets.
Additionally, Chen et al. \cite{AMD_M3} proposed the M3 framework (Multi-modal, Multi-task, Multi-attention), which introduces self-attention and cross-modal attention mechanisms, enables automated detection of RPD. RPD, an important marker of AMD progression, is difficult to identify in traditional CFP, but this method, combined with FAF image information, enables even primary healthcare institutions with only CFP capabilities to achieve high-precision RPD identification.

\subsubsection{Glaucoma Diagnosis}
Glaucoma is a group of eye diseases characterized by optic nerve damage and visual field defects, and it is the leading cause of irreversible blindness worldwide. Its pathogenesis is complex, and early symptoms are often subtle, leading to diagnosis typically occurring only when the disease has progressed to the moderate-to-late stages. Traditional diagnostic methods rely on optic disc structure analysis, visual field testing, and intraocular pressure measurement. However, due to significant anatomical variability among individuals and the subjective nature of visual field testing, misdiagnosis and missed diagnoses are common in clinical practice. Additionally, long-term follow-up management of glaucoma patients faces challenges in monitoring changes in both structural and functional indicators \cite{Glau2023_GRAPE, Glau2023_Pseudo_Supervisor}.

In response to the aforementioned challenges, researchers have gradually explored new approaches combining multimodal imaging data with deep learning. Mehta et al. \cite{Glau2021_DenseNet} developed a multimodal machine learning model integrating OCT, CFP images, and demographic and clinical data, thereby improving the accuracy of early glaucoma detection. Li et al. \cite{Glau2022_} compared three different multimodal information fusion strategies (early fusion, mid-fusion, and hierarchical fusion), finding that the hierarchical fusion strategy performed best on the GAMMA dataset, effectively addressing the complementarity and correlation between modalities.
Zhou et al. \cite{Glau2023_MM_RAF} proposed the MM-RAF framework based on the Transformer architecture, incorporating bilateral contrast alignment (BCA), multi-instance learning representation (MILR), and hierarchical attention fusion (HAF) modules to overcome the insufficient interaction of spatial information between different modalities, significantly enhancing the robustness of glaucoma identification. Meanwhile, the GAMMA Challenge organized by Wu et al. \cite{Glau2023_GAMMA} has driven the development of glaucoma grading algorithms, with the 3D-DEN dual-branch architecture and multi-model integration strategy adopted by participating teams demonstrating excellent performance on multimodal data.

\subsubsection{Diabetic Retinopathy Diagnosis}
DR is one of the most common microvascular complications of diabetes, and severe cases can lead to vision loss or even blindness. Its pathological process includes retinal microvascular leakage and neovascularization, requiring regular screening through fundus photography, OCTA, and other methods. However, the grading standards for DR are complex, manual image interpretation is inefficient, and existing automated screening systems have limitations in multi-modal data processing, particularly in terms of generalization ability on small datasets.

Hua et al. \cite{DR2020_TFA_Net} proposed the TFA-Net model, which integrates fundus images with SS-OCTA data. By utilizing weight-sharing convolutional kernels and reverse cross-attention (RCA) streams to enhance feature representation, the model significantly improves DR classification performance on small datasets. Song et al. \cite{DR2021_MMIB_Net} developed the MMIB-Net model based on the information bottleneck theory, which extracts features relevant to the classification task while removing redundant information, enhancing the model's discriminative ability.
Alvaro et al. \cite{DR2022_MIE} proposed a new method based on multimodal image encoding pre-training (MIE), utilizing unlabeled retinal image pairs for cross-modal feature learning, enhancing the model's generalization ability under unsupervised conditions. Mostafa et al. \cite{DR2023_Mostafa} further optimized the classification performance of UWF-CFP and OCTA image fusion by combining ResNet50 and 3D-ResNet50 modeling with SE blocks and Manifold Mixup technology.

\subsubsection{Multi-disease Diagnosis}
There are many types of ophthalmic diseases, and they often coexist or occur concurrently in the same patient. Diagnosing a single disease is no longer sufficient to meet the needs of complex clinical scenarios. Most existing studies focus on single disease identification and lack support for multi-disease joint diagnosis.\cite{Multi2021_MM_MIL, Multi2021_Kang, Multi2022_Uni4Eye, Multi2024_Uni4eye_plusplus} In addition, clinical data usually exhibits a long-tail distribution (i.e., there are few samples of certain rare diseases). How to achieve efficient multi-disease classification with limited labeled data has become a major challenge.

Li et al. \cite{Multi2020_LiXiaomengi} proposed a multi-modal feature learning method based on self-supervised learning, which enables the model to capture cross-modal semantic shared information by synthesizing FFA images and jointly training them with CFP. He et al. \cite{Multi2021_MSAN} designed a modality-specific attention network (MSAN) that extracts key features from CFP and OCT images using multi-scale attention modules and region-guided attention modules, respectively, enhancing the accuracy of multi-disease classification.
Ou et al. \cite{Multi2021_M2LC_Net} proposed the M2LC-Net, which adopts a ResNet18-CBAM structure combined with class balance loss (CBL) and a two-stage training strategy, effectively mitigating classification bias caused by long-tail distributions. Zou et al. \cite{Multi2024_EyeMost_plus} developed the EyeMoSt+ model, which introduces a confidence-aware mechanism, combining uncertainty modeling with the Mixed Student t-distribution (MoSt) for multimodal fusion, enhancing the model's robustness and generalization capabilities under noisy conditions.

\subsection{Multimodal Ophthalmic Data Generation and Augmentation}
In the field of medical image analysis, generation technology is gradually becoming an important tool to assist clinical diagnosis.Especially in fundus image analysis, its highly specialized anatomical structure and rich pathological information have become important research objects for both Image-to-Text Generation and Image-to-Image Translation tasks. Such tasks (as shown in Table \ref{Generation}) not only help compensate for the lack of information in single-modality images but also provide alternative diagnostic tools for healthcare organizations that lack modality-specific data.

\begin{table*}[t]
\centering
\fontsize{7}{9}\selectfont
\caption{Summary of multimodal approaches for ophthalmic image-to-text generation and image translation.}
\label{Generation}
\begin{tabularx}{\textwidth}{%
  >{\RaggedRight\arraybackslash}p{0.4cm}  
  >{\RaggedRight\arraybackslash}p{2.0cm}  
  >{\RaggedRight\arraybackslash}p{2.8cm}  
  >{\RaggedRight\arraybackslash}p{3.0cm}  
  >{\RaggedRight\arraybackslash}p{2.4cm}  
  >{\RaggedRight\arraybackslash}p{3.0cm}  
}
\toprule
\textbf{Year} & \textbf{Study} & \textbf{Approach} & \textbf{Application} & \textbf{Modality} & \textbf{Dataset} \\
\toprule
\multicolumn{6}{@{}p{\textwidth}@{}}{\quad Image to Text}\\
\midrule
2021 &Huang et al. \cite{ger1} &GPT-2, VGG, Transformer &Retinal Report Generation &CFP, FFA, Description &DeepEyeNet \\
2021 &Huang et al. \cite{ger11} &MobileNet/VGG/Inception, LSTM, CNN &Retinal Report Generation &CFP, FFA, Description &DeepEyeNet \\
2022 &Li et al. \cite{ger8} &CNN, Transformer &Retinal Report Generation &FFA, Report &FFA-IR \\
2022 &Huang et al. \cite{ger5} &CNN, LSTM, Transformer &Retinal Report Generation &CFP, FFA, Description &DeepEyeNet \\
2022 &Sun et al. \cite{ger7} &CNN, Bert, Transformer &Retinal Report Generation &FFA, Reports &FFA-IR \\
2023 &Bu et al. \cite{ger3} &CNN, Transformer &Mdedical Report Generation &X-Ray, FFA, Report &IU X-Ray, MIMIC-CXR, FFA-IR \\
2024 &Shaik et. al \cite{ger2} &VGG, Transformer &Retinal Report Generation &CFP, FFA, Description &DeepEyeNet \\
2024 &Shaik et. al \cite{ger4} &EfficientNet, Transformer &Retinal Report Generation &CFP, FFA, Description &DeepEyeNet \\

\midrule
\multicolumn{6}{@{}p{\textwidth}@{}}{\quad Image to Image}\\
\midrule
2020 &Tavakkoli et al. \cite{ger12} &CNN, cGAN &CFP to FFA translation &CFP, FFA &Isfahan MISP \\
2021 &Kamran et al. \cite{ger19} &CNN, GAN, PatchGAN &CFP to FFA translation &CFP, FFA &Isfahan MISP \\
2022 &Chen et al. \cite{ger18} &CNN, cGAN &CFP to FFA translation &CFP, FFA &Private \\
2023 &Huang et al. \cite{ger15} &CNN, R-cGAN, PatchGAN &CFP to FFA translation &CFP, FFA &Private \\
2023 &Shi et al. \cite{ger20} &GAN &CFP to FFA translation &CFP, FFA &Private \\
2023 &Kang et al. \cite{ger21} &CycleGAN &CFP to FFA translation &CFP, FFA &Private \\
2024 &Zhao et al. \cite{ger16} &CNN, Attention, GAN &CFP to FFA translation &CFP, FFA &Isfahan MISP \\
2024 &Kamran et al. \cite{ger14} &CNN, cGAN &CFP to FFA translation &CFP, FFA &Isfahan MISP \\
2024 &Wang et al. \cite{ger17} &Diffusion, GAN &CFP to FFA translation &CFP, FFA &MPOS \\
2024 &Chen et al. \cite{ger23} &GAN &CFP to ICGA translation &CFP, ICGA &Private \\
2024 &Li et al. \cite{ger24} &GAN &OCT to OCTA Translation &OCT, OCTA &OCTA-500 \\
2025 &Yuan et al. \cite{ger22} &ResNet, GAN, PatchGAN &CFP to FFA translation &CFP, FFA &Isfahan MISP \\
\bottomrule
\end{tabularx}
\end{table*}

\subsubsection{Image-to-Text Generation}
The goal of medical image-to-text generation is to transform the content of medical images into accurate and interpretable text reports. Unlike general image captioning, medical description generation must capture domain-specific semantics and support clinical decision-making, thereby requiring deeper alignment between medical image and medical knowledge.

Existing approaches typically follow a progressive modeling strategy that gradually enhances the multi-modal semantic representation.First, encoding clinical keyword to improve the semantic quality of the contextualized representations and serve as strong priors for downstream generation tasks. Building upon this case, subsequent efforts focus on refining visual feature extraction, aiming to obtain more semantically rich and clinically relevant visual representations. These enhanced visual features are then aligned with keyword embeddings through multimodal semantic interaction strategies, enabling capturing the semantic relationships between modalities.Finally, to bridge the gap between model prediction and clinical expertise, medical knowledge infusion mechanisms incorporate external knowledge to enhance the reliability of generated reports.

In the medical image-text generation task, the \textbf{keywords context encoder} plays a central role, as the effective representation of keywords has always been one of the key research focuses. Since static word embeddings (e.g. Word2Vec, GloVe) are unable to distinguish polysemous words or capture contextual semantic changes, some studies have made attempts to improve the quality of keyword encoding. Huang et al. \cite{ger1} uses GPT-2 to generate context-sensitive word representations, which solves the problem that static word embeddings cannot distinguish polysemous words, and can better take into account the contextual semantics of the text (keywords); Gated Contextual Attention Net (GCA-Net) \cite{ger2} encodes clinical keywords using an inclusion embedding layer and an attention module to generate context-aware keyword embedding representations to express keywords. These methods improve the quality of encoding textual (keyword) information, which in turn enhances the generation of multimodal medical image descriptions. 

Although contextual information plays a significant role in generating medical descriptions, the \textbf{refinement of visual feature extraction} remains essential for high-quality outcomes. For instance, in multi-view or multi-slice scenarios, conventional approaches tend to directly aggregate features across slices, potentially discarding critical diagnostic cues. In contrast, ReFuGen \cite{ger3} employs a pre-trained ResNet-101 to extract semantic features of the image slices, which are further refined by the Adaptive Slice Selection (ASS) and Spatial Refinement (SPR) branches to highlight key slices and integrate global information, thereby preserving both slice-specific and complementary diagnostic information. However, there is a issue with using pre-trained convolutional neural networks to extract visual features. Since each feature map is given the same level of priority, most features may not be salient enough to help the model make accurate decisions. Therefore, GCA-Net and M3 Transformer \cite{ger4} introduce an attention mechanism to prioritize significant features. They design a Gated Contextual Attention module, which enhances both local and global contextual information through global attention pooling, channel dependency capturing, and adaptive feature modulation, thereby obtaining more discriminative visual features. 

With enhanced capabilities in extracting contextual and visual features, research has further advanced into \textbf{multimodal semantic interaction}, aiming to model richer cross-modal relationships. Some studies are exploring how to utilize expert-defined keywords to direct image attention. Shaik et al. \cite{ger2, ger4}, Huang et al. \cite{ger5} and Wu et al. \cite{ger6} integrates visual features and keyword embedding by employing the Transformer architecture and utilizes a multi-head self-attention mechanism to capture the visual features and the semantic relationships between visual features and keywords using a multi-head self-attention mechanism to provide attention-weighted image-keyword information. And Sun et al. \cite{ger7} projects the visual features of fundus images into the semantic space and then concatenate them with their corresponding word embeddings as fused cross-modal information feature.

Beyond aligning visual and textual modalities, recent studies have increasingly explored \textbf{medical knowledge infusion mechanisms} to enhance models’ reasoning capabilities and generalization performance. CGT \cite{ger8} proposes a method that incorporates clinical relation triples into visual features as prior knowledge to guide the decoding process. This approach utilize a Transformer-based cross-modal clinical graph network to integrate the NLP-Built knowledge graph with image features, enabling the generated reports to better align with medical logic.

Along with above approaches,there are also studies exploring other innovations. Wen et al. \cite{ger10} introduce causal chains to pass symptom information to the pathology branch, so that the generated pathology descriptions are more consistent with medical logic; Moreover, Huang et al. \cite{ger11} utilize CAM technology to visualize the region of interest of the model and validate its consistency with the ophthalmologist, and so on these studies are pushing the further development of iamge-to-text models in terms of reliability and interpretability.

Looking ahead, as more high-quality medical image–text pairs become available and large-scale pre-trained models continue to improve, image-to-text generation from medical images is expected to play an increasingly important role in clinical practice.

\subsubsection{Image Translation}
In ophthalmologic image translation, \textbf{GAN-based approaches} have gained significant traction, particularly in cross-modality generation tasks such as converting CFP to FFA. However, given the unique structural and texture characteristics of ophthalmologic images, traditional GAN architectures often face challenges such as mode collapse, loss of anatomical consistency, or failure to preserve pathological details.

The fundus image translation tasks has undergone an iterative evolution of different technical approaches, mainly focusing on solving the problems of image quality, detail retention and model stability.To address these issues, many studies have proposed improvements, including the advances in the GAN architecture itself, as well as optimizations of loss functions. It reflects a natural evolution from model capability (via architectural design), to training effectiveness (via loss optimization). Moreover, tasks have expanded from the common CFA-to-FFA to other tasks, such as CFA-to-ICGA, reflecting the growing breadth and application scope of GAN-based ophthalmic models.The evolution offering a comprehensive perspective on how GAN-based methods are advancing ophthalmologic image translation.

To enhance the quality and detail fidelity of generated images in medical image translation tasks, recent research has focused on both architectural innovations and the optimization of generators and discriminators in \textbf{GANs}. Multi-scale and multi-stage generator-discriminator frameworks, along with attention mechanisms, have emerged as key approaches to improving translation performance.

From an architectural perspective, several studies have proposed multi-scale generator-discriminator structures to effectively model both global structures and fine-grained details. For example, models such as Novel cGAN, Attention2AngioGAN, FA4SANS-GAN and LA-GAN \cite{ger12,ger13,ger14,ger15} introduce a hierarchical design where the coarse-grained generator constructs the large-scale structures of the FFA image, such as macula, optic disc, color, and brightness, whereas fine-grained generator is used to generate detailed information like small vessels, hemorrhages, and exudates. Correspondingly, The coarse-grained discriminator works on half-resolution image pairs, while the fine-grained one operates at full resolution, preserving global coherence while enriching fine details.

Besides architectural improvements, innovations within the generator and discriminator designs are crucial for improving image fidelity and maintaining training stability. For instance, Novel cGAN and Attention2AngioGAN integrate global and local features by fusing residual block outputs with fine-grained convolution layers in the generator, facilitating shared feature space learning. Attention2AngioGAN further employs attention blocks to fuse deep and shallow features, preserving the spatial information. Unlike Attention2AngioGAN’s motivation, DCLAGAN \cite{ger16} employs a different attention module.It incorporates a Coordinate Attention (CA) module into the generator's encoder to better capture spatial positions, and integrates Class Activation Mapping (CAM) modules in both the decoder and discriminator to emphasize critical regions in the images. Additionally, for specific translation tasks, only minor modifications within the generator are needed to adapt it to the new task. In category-specific tasks, Dynamic Diffusion-guided GAN \cite{ger17} embeds category priors into the generator, providing semantic guidance and improving class-specific synthesis. And for tasks requiring the generation of multi-phase FFA images, like SPFSAT \cite{ger18}, for instance, employs a shared encoder and multiple decoders in generator to simultaneously generate multiple phases of an FFA image, enhancing the ability to capture phase-wise structural consistency through unified feature learning.

Correspondingly, discriminator innovations also play a key role, with some studies making novel contributions in this area. Dynamic Diffusion-guided GAN introduces a diffusion process by injecting noise into both real and generated images during training, which expands the distribution support and mitigates mode collapse. On the other hand, VTGAN \cite{ger19} designs a Markovian discriminator based on the Visual Transformer, leveraging its self-attention mechanism to capture long-range dependencies while preserving fine local features.
These innovative designs within the generator and discriminator, significantly, improve the the flexibility of GAN-based translation methods in handling different task and the performance of generating high-quality translated ophthalmologic images.

\textbf{Optimization of Loss Function} has seen researchers working on refining the loss function to achieve better detail quality and more stable training in image generation tasks. For example, Shi et al. \cite{ger20} introduced a gradient variance loss on top of the pix2pixHD model introduces a gradient variance loss to better capture detailed information such as blood vessels and textures by focusing on the high-frequency components of the image; CycleEBGAN \cite{ger21}, on the other hand, avoids generating image clustering problems as well as pattern collapse by introducing an energy function and redefining the adversarial loss as the distance between the simulated image and the real image in the discriminator output space (MSE). While Multiple-ResNet GAN \cite{ger22} uses least-square GANS(LSGAN) to optimize the training process and make the GAN-based training more stable.\\

Although most of the work on translation tasks on fundus images has been performed on the task of CFP to FFA, there is no lack of research on other modal translations, for example, Chen et al. \cite{ger23} investigated the task of CFP to ICGA image translation, the image translation was performed by using pix2pixHD, the generation of high-frequency details was enhanced by introducing the gradient variance loss, and the diagnostic value of the generated images was verified in the age-related macular degeneration (AMD) classification task; TransPro \cite{ger24} is targeted at the OCT to OCTA translation task. A 3D convolutional neural network is used to process 3D OCT volume to 3D OCTA translation and a 2D generative network is used to process 2D projection maps, supplementing the contextual information of the 3D generative network with a heuristic contextual guidance module (HCG) for alignment. A vascular promotion guidance module (VPG) is also introduced to improve the generation accuracy of the vascular regions.

With the introduction of the above methods, we can see that the CFP to FFA translation task has made significant progress in terms of improving image quality, detail retention, and the stability of model training. The innovations at each stage are closely centered on how to better simulate real medical images and how to provide more valuable information for clinical diagnosis, continuously advancing the field of medical image translation.

\section{Ophthalmic Multimodal Foundation Models}
In contrast to the previous chapter’s task-specific discussion centered on specific clinical applications, this chapter focuses on the \textbf{model architectures} themselves, exploring the evolution and application of multimodal foundation models in ophthalmology from a technical design perspective. As shown in Table \ref{foundation}, to systematically present the development trajectory of these models, this section categorizes them into four main types: masked images modeling base (MIM-base) vision models \cite{MIM}, contrastive language–image pretraining style (CLIP-style) models \cite{CLIP}, multimodal large language models (MLLMs) \cite{MLLM}, and Fine-tuned Foundation Models \cite{fineturning}.

\begin{figure*}[t]
\centering
\includegraphics[width=1\textwidth]{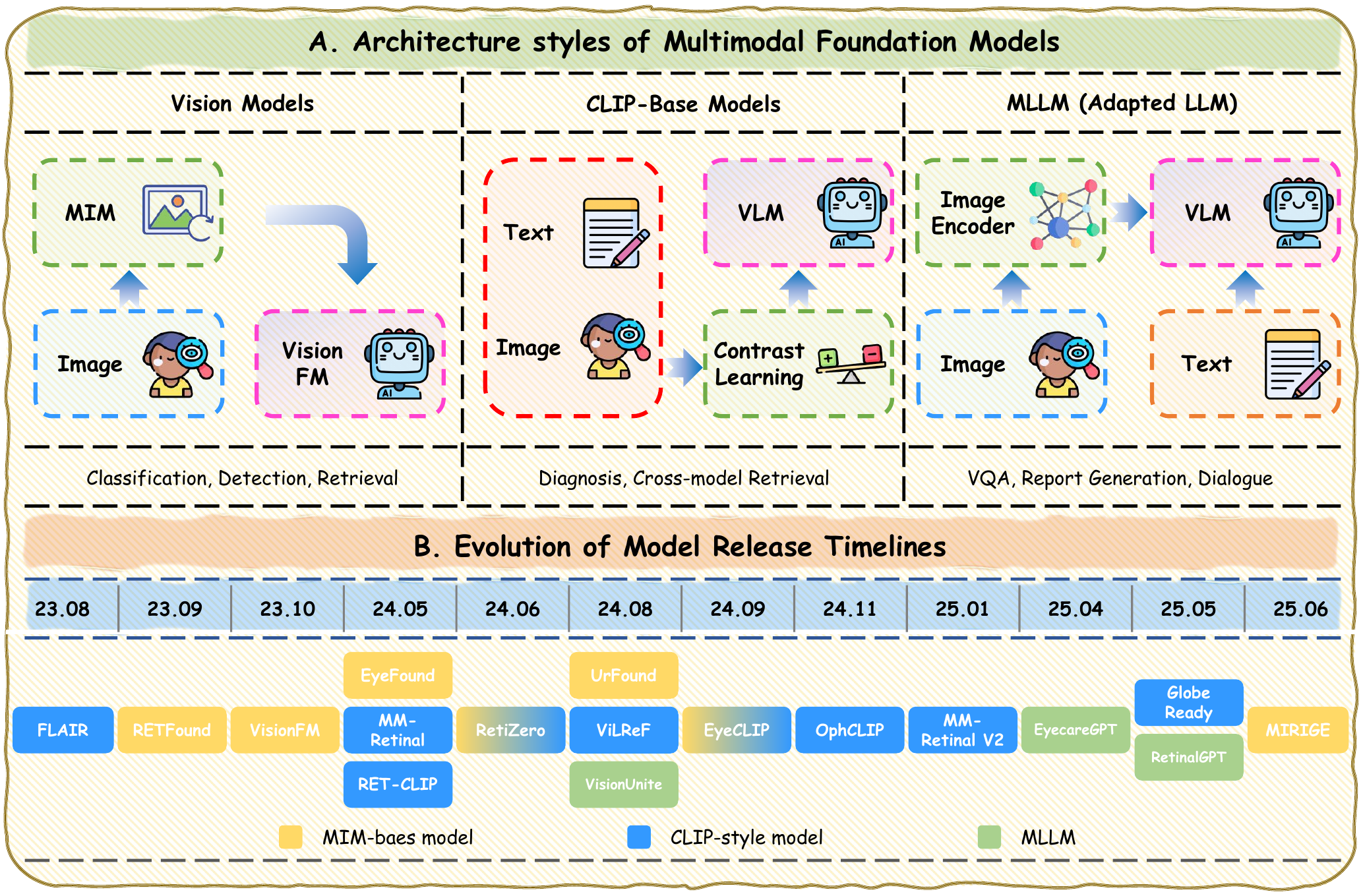}
\caption{Overview and Evolution of Foundation Models in Ophthalmic Multimodal Learning}
\label{fig4}
\end{figure*}

Specifically, \textbf{MIM-based vision models} that mainly rely on image reconstruction techniques use various ophthalmic imaging modalities as inputs and demonstrate remarkable generalization capabilities. \textbf{CLIP-style models} employ contrastive learning to effectively align ophthalmic images with clinical text, which enhances cross-modal understanding and interpretability. Among these, \textbf{MLLMs} are currently the most generalized models because they leverage advanced conversational and multitask processing abilities to advance ophthalmic AI toward higher levels of cognition and reasoning. An overview of these three categories of foundation models, together with their chronological development in ophthalmic research, is illustrated in Figure~\ref{fig4}.

In addition to these, \textbf{Fine-tuned Foundation Models} represent a growing class of approaches. These models adapt general-purpose pretrained models through architectural modifications and task-specific supervision. By doing so, they offer a practical balance as they retain the expressive power of large-scale pretraining while improving applicability, efficiency, and robustness for particular ophthalmic tasks. The following sections will provide detailed descriptions of the technical features and representative applications of these four categories of models.

\subsection{MIM-base Vision Models}

Ophthalmic imaging data often lack sufficient annotations due to the high cost and specialized expertise required for labeling. To overcome this limitation, self-supervised learning has emerged as an effective approach for training foundational models on large-scale unlabeled datasets. Although RETFound \cite{intro14} is not a multimodal method, it represents a milestone in ophthalmic foundation modeling. Focusing on CFP and OCT modalities, RETFound was pretrained on 1.6 million unlabeled retinal images using masked autoencoding, pioneering the application of masked image modeling (MIM) in this field. It has demonstrated strong performance across various downstream tasks, including the diagnosis of diabetic retinopathy and glaucoma, as well as the prediction of systemic conditions like heart failure, establishing itself as a key benchmark for subsequent multimodal models.

Then, VisionFM \cite{visionfm} introduces a more sophisticated multimodal framework by employing eight independent encoders, each adapted to a specific modality for self-supervised pretraining. This design enables modality-specific feature extraction while maintaining a unified framework. In comparison, EyeFound \cite{eyefound} proposes a unified approach that utilizes a single encoder for all 11 modalities. By pretraining with masked autoencoding on 2.78 million images collected from 227 hospitals, EyeFound captures shared representations across modalities and outperforms RETFound in systemic disease prediction tasks. This highlights the advantages of joint multimodal representation learning

A recent trend in MIM-based approaches is reducing reliance on large-scale annotated data. UrFound \cite{urfound} makes a notable breakthrough by incorporating textual supervision into the MIM framework. Similar to EyeFound, it employs a shared encoder for CFP and OCT images along with two decoders, one for masked image modeling (MIM) and the other for masked language modeling (MLM). Through cross-modal pretraining enabled by cross-attention mechanisms, UrFound strengthens visual-textual alignment. Remarkably, it surpasses RETFound on both CFP and OCT tasks using only 180k pretraining images. In contrast, MIRAGE \cite{mirage} focuses on enhancing inter-modality alignment. By leveraging strictly paired OCT and SLO images acquired from the same patient during the same scanning session, MIRAGE employs a MultiMAE framework for pretraining to explicitly align the two modalities.Experiments demonstrate that MIRAGE, trained on 260k image pairs, outperforms RETFound across multiple benchmarks

\subsection{CLIP-style Models}
Text-image models typically adopt two predominant architectures: CLIP-style models that align different modalities, and multimodal large language models that integrate cross-modal understanding. The CLIP-style model demonstrates the power of multimodal learning by aligning text and images via contrastive learning. This framework offers a promising approach for ophthalmology, enabling effective integration of imaging data (e.g., fundus, OCT) with clinical reports. Such aligned multimodal representations could improve disease detection, enhance interpretability, and even facilitate personalized treatment recommendations, ultimately bridging the gap between imaging findings and clinical decision-making.

FLAIR \cite{flair} marked an early attempt to apply the CLIP architecture to ophthalmology. By converting categorical disease labels into template-based textual descriptions, it enriched the semantic context during training and improved generalization in zero-shot scenarios. However, its performance remained limited due to the lack of real-world clinical data. RET-CLIP \cite{retclip} advanced this direction by leveraging real image–text pairs from diagnostic reports of 193,865 patients. A key innovation was its tripartite optimization strategy across left eye, right eye, and patient-level features, which better captured the bilateral nature of ocular assessments and significantly enhanced generalization across diverse retinal diseases.

Based on the above model, MM-Retinal \cite{mmretinal} introduced a mixed supervision strategy that combined publicly available image label datasets with a newly constructed multimodal ophthalmology corpus. This corpus consists of high-resolution CFP FFA and OCT images paired with long-form expert-written reports. The method proposed a novel knowledge transfer module that refines representations from public data through retrieval-based distillation using its fine-grained corpus. This approach improves zero-shot and few-shot transfer capabilities across different domains. Compared to its predecessor, MM-Retinalv2 \cite{mmretinalv2} expanded the dataset scale by three times and introduced a dual path knowledge injection mechanism. In addition to the semantic level alignment present in the first version, it incorporated appearance level features using vector quantization, which allows the model to capture fine-grained lesion details. With the addition of a text pretraining stage based on a carefully curated ophthalmology corpus, the resulting model, KeepFIT v2, demonstrated superior performance. It outperformed several large-scale models on tasks involving CFP, FFA, and OCT modalities.

RetiZero \cite{retizero} differs from previous methods by integrating masked autoencoders with CLIP and introducing a Dirichlet reparameterization framework that explicitly models uncertainty within the visual language embedding space. Rather than producing fixed embeddings, this design generates distributional representations that quantify prediction confidence. This uncertainty-aware modeling greatly improves robustness in long-tailed and open-set conditions and enables reliable zero-shot recognition across more than four hundred fundus disease categories. EyeCLIP \cite{eyeclip} also uses a masked autoencoder and CLIP hybrid architecture while focusing on multimodal alignment. It employs a three-stage pretraining pipeline that includes masked self-supervised reconstruction, cross-modal image contrastive learning, and image-text contrastive learning. The unified encoder effectively captures complementary information across eleven imaging modalities and excels in few-shot and cross-modal transfer scenarios.

Although CLIP-based models have shown promise, they encounter obstacles such as false negative pairs in contrastive learning. ViLRef \cite{vilref} addresses this issue with a knowledge guided negative suppression mechanism that uses expert-annotated clinical labels as guidance. The model calculates semantic similarity between sample pairs and adjusts the contrastive loss dynamically so that semantically similar examples are not penalized. The use of a momentum encoder along with a queue based memory increases the diversity of training samples and results in improved classification and lesion localization on various benchmarks. For practical deployment, GlobeReady \cite{ophclip} responds to the need for model adaptation by applying retrieval augmented generation to bring in relevant local data features automatically. This approach removes computational burdens and the need for expert intervention. In addition, it applies Bayesian uncertainty estimation to deliver predictions with risk awareness, which helps to lower both misdiagnosis and missed diagnoses. The method also demonstrates strong generalization in a variety of clinical environments.

Expanding the scope beyond fundus images, OphCLIP \cite{ophclip} targeted surgical video understanding by introducing a hierarchical video-language alignment framework. It aligned short clips with surgical narrations and full videos with titles, leveraging a multi-granular contrastive loss design. An innovative memory bank mechanism was used to dynamically retrieve relevant "silent" videos to enhance supervision, significantly improving zero-shot performance in the surgical phase and instrument recognition.

\subsection{Multimodal Large Language Models}
Although contrastive learning frameworks have demonstrated strong capabilities in aligning visual and textual modalities, they often fall short in supporting complex reasoning and generation tasks. Recently, MLLMs, which integrate visual encoders with large-scale language models, have emerged as a promising paradigm to overcome these limitations. By combining the perceptual strengths of vision models with the generative and inferential capacities of LLMs, MLLMs enable more comprehensive multimodal understanding and task execution. In the context of ophthalmic disease diagnosis and clinical decision-making, such models exhibit substantial potential and are poised to play a transformative role.

VisionUnite \cite{visionunite} represents one of the earliest efforts to develop a multimodal large language model specifically designed for ophthalmology. It integrates a CLIP-based vision encoder with an LLaMA-based language model and enhances performance through a vision adapter and projector that enable fine-grained alignment between visual embeddings and textual prompts. The model is trained on MMFundus, which is the largest multimodal retinal dataset available. VisionUnite supports multi-round interaction, open-ended diagnosis across various retinal conditions, and generation of clinically grounded reports. By applying lesion level classification supervision and contrastive alignment it achieves strong performance in diagnosis accuracy and interpretability. This model outperforms GPT 4V and Gemini Pro on several benchmark tasks.

EyecareGPT \cite{eyecaregpt}, on the other hand, targets general-purpose ophthalmic understanding across a wide range of imaging modalities. Built upon the comprehensive Eyecare-100K dataset covering 8 imaging types and over 100 diseases, it introduces a Layerwise Dense Connector (LDC) to fuse multiscale visual features and a resolution-adaptive module (AnyRes) to handle heterogeneous clinical image inputs. These innovations allow EyecareGPT to capture fine-grained local structures, significantly improving performance on tasks such as closed/open-ended QA and report generation. With higher fidelity in anatomical and pathological comprehension, it surpasses prior models in both structured and free-form evaluations.

\subsection{Fine-tuned Foundation Models for Ophthalmology}
While foundation models demonstrate strong generalization across a wide range of tasks, their deployment typically requires substantial computational resources and large-scale data. In contrast, fine-tuning offers a more cost-effective alternative. MLLMs represent a primary outcome of such strategies, where a dedicated visual encoder is trained and its extracted features are embedded into a language model to enable cross-modal integration and task execution.

This approach finds its earliest ophthalmic implementation in OphGLM \cite{ophglm}, which strategically combines a pretrained fundus image encoder with a large language model. Targeted fine-tuning on the Chinese-language Fundus Tuning CN dataset optimizes retinal disease classification while maintaining computational efficiency. It surpasses open source baseline models in diagnostic accuracy without the need for full model retraining. Extending this paradigm to fluorescein angiography, FFA GPT \cite{ffagpt} achieves even greater parameter efficiency. Its self supervised pretrained visual module undergoes lightweight adaptation before integration with LLaMA 2 and enables both automated report generation and interactive question answering. These cases collectively demonstrate that task specific excellence in ophthalmology can be achieved through minimal architectural adjustments and focused domain data.

Extending these task-specific innovations, RetinaVLM \cite{retinalvlm} and DeepDR-LLM \cite{deepdr} place greater emphasis on multimodal alignment and disease-specific management. RetinaVLM employs complex pretraining strategies to optimize the fusion of visual and textual information, thereby improving generalizability across a wide range of ophthalmic disorders. DeepDR-LLM, with a focus on DR, integrates Transformer architectures with LLMs to accomplish image quality assessment, lesion segmentation, and disease grading. Its clinical decision-support performance has been shown to be on par with that of junior ophthalmologists.

Furthermore, RetinalGPT \cite{retinalgpt} advances the field by demonstrating exceptional versatility through the integration of retinal image analysis with structured vascular fractal features such as branching angles and fractal dimensions extracted using AutoMorph and RBAD tools. Its training proceeds in two stages. First, feature alignment is conducted using PMC-600K and RCA data. Second, hybrid instruction tuning is performed with RCT and medical question-answering data. This approach achieves superior performance across eight retinal datasets and surpasses general-domain multimodal large language models like GPT-4 in tasks including lesion localization and generation of clinically interpretable outputs.

While research on fine-tuning MIM models is relatively scarce, FMUE \cite{fmue} has made breakthroughs in this field: built on RETFound's encoder, it is fine-tuned with only 102,468 OCT images but performs better in multi-disease classification tasks, achieving an F1 score of 95.74\% on the internal test set, surpassing RETFound's 93.34\%. More innovatively, it introduces uncertainty classification based on the Dirichlet distribution, which can output prediction reliability scores, effectively enhancing the safety of clinical applications.

\begin{table*}[t]
\centering
\fontsize{7}{9}\selectfont
\caption{Overview of multimodal foundation models in ophthalmology.}
\label{foundation}
\begin{tabularx}{\textwidth}{%
  >{\RaggedRight\arraybackslash}p{0.4cm}  
  >{\RaggedRight\arraybackslash}p{2.0cm}  
  >{\centering\arraybackslash}p{3.4cm}  
  >{\centering\arraybackslash}p{2.2cm}  
  >{\centering\arraybackslash}p{3.6cm}  
  >{\centering\arraybackslash}p{2.0cm}  
}
\toprule
\textbf{Year} & \textbf{FM in Study} & \textbf{Modality} & \textbf{Pretrain Samples} & \textbf{Application} & \textbf{Venue}\\ 
\midrule
\multicolumn{6}{@{}p{\textwidth}@{}}{\quad Multimodal Vision Models}\\
\midrule
2022    &RETFound \cite{intro14}   &CFP, OCT & 904,170 + 736,442 &Diagnosis & Nature\\
2023    &VisionFM \cite{visionfm}   &CFP, OCT, FFA, Slit-Lamp, MRI, UBM,B-Scan, External & 3,559,418 in total  &Diagnosis, Segmentation & NEJM AI \\
2024    &EyeFound \cite{eyefound}   &CFP, OCT, FFA, Slit-Lamp, ICGA, OUS, FAF, Specular, External, Topo, RetCam & 2,777,593 in total &Diagnosis, VQA & arXiv\\
2024    &UrFound \cite{urfound}    &CFP, OCT &102,468 + 83,484 & Diagnosis & MICCAI\\
2025    &MIRAGE \cite{mirage}     &OCT, SLO   & 261,184 pairs & Diagnosis, Segmentation   & arXiv\\
\midrule
\multicolumn{6}{@{}p{\textwidth}@{}}{\quad CLIP-Base Models}\\
\midrule
2023    &FLAIR \cite{flair}  &CFP, Text & 284,600 pairs &Diagnosis  & Med Image Anal\\
2024    &RET-CLIP \cite{retclip}   &CFP, Text & 193,865 triplets &Diagnosis & MICCAI\\
2024    &KeepFIT \cite{mmretinal} &CFP, FFA, Text &(280,517 + 1,050,531)pairs    &Diagnosis, Report Generation   &MICCAI\\
2024    &RetiZero \cite{retizero}   &CFP, Text & 341,896 pairs &Diagnosis, Retrieve & Nat. Commun.\\
2024    &ViLReF \cite{vilref}     &CFP, Text  &451,956 pairs  &Diagnosis, Segmentation    &arXiv\\
2024    &EyeCLIP \cite{eyeclip}    &(CFP, OCT, FFA, Slit-Lamp, ICGA, OUS, FAF, Specular, External, Topo, RetCam), Text & 2,777,593 pairs &Diagnosis, VQA, Retrieve & npj Digit. Med.\\
2024    &OphCLIP \cite{ophclip}    &Video, Text    &375,198 pairs  &Detection  &arXiv\\
2025    &KeepFIT-v2 \cite{mmretinalv2}  &CFP, FFA, OCT, Text    &(276,720 + 1,053,703 + 186,502) pairs& Diagnosis, Report Generation    &arXiv\\
2025    &GlobeReady \cite{globeready} &CFP, OCT, Text &(25,656,951 + 13,207,324) pairs    &Diagnosis  &arXiv\\
\midrule
\multicolumn{6}{@{}p{\textwidth}@{}}{\quad Multimodal Large Language Models}\\
\midrule
2024    &VisionUnite \cite{visionunite}    &CFP, Text & - &Diagnosis, VQA, Report Generation & arXiv\\
2025    &EyecareGPT \cite{eyecaregpt}     &CFP, OCT, FFA, Slit-Lamp, ICGA, UBM, FS, CT, VQA  &58,485 in total + 102,000    &VQA, Report Generation, Detection &arXiv\\
\midrule
\multicolumn{6}{@{}p{\textwidth}@{}}{\quad Fine-tuned Foundation Models}\\
\midrule
2023    &OphGLM \cite{ophglm}     &CFP, VQA & 106,663 + over 20,000 &Diagnosis, VQA, Report Generation & Artif Intell Med\\
2023    &FFA-GPT \cite{ffagpt}    &FFA, Text & 421,916 pairs &VQA, Report Generation & npj Digit. Med.\\
2024    &RetinaVLM \cite{retinalvlm}  &OCT, VQA & 44,733 + 479,710 &Diagnosis, VQA, Report Generation & arXiv\\
2024    &DeepDR-LLM \cite{deepdr} &CFP, Recs. & 1,247,135 + 371,763 &Report Generation  & Nat Med\\
2025    &RetinalGPT \cite{retinalgpt} &CFP, Text &736,000 pairs   &Diagnosis  &arXiv\\
2024    &FMUE \cite{fmue}       &OCT       &102,468     &Diagnosis &CELL REP MED\\
\bottomrule
\end{tabularx}
\label{tab:fm_models}
\end{table*}

\section{Overview of Ophthalmic Imaging Datasets}
The development and evaluation of multimodal deep learning models in ophthalmology heavily rely on access to high-quality, diverse datasets. This review compiles a wide range of commonly used datasets that span various imaging modalities and ophthalmic conditions, providing crucial support for research on disease diagnosis, lesion segmentation, and data generation. These datasets are categorized into four major areas that reflect key directions in ophthalmic AI research.

Datasets related to age-related macular degeneration (AMD) (Table \ref{amd-dataset}), such as AREDS, the Duke series, and OCTiD, offer OCT and color fundus photography (CFP) images for studying AMD progression and subtypes. In the domain of diabetic retinopathy (DR) (Table \ref{dr-dataset}), datasets like IDRiD, Messidor, and DDR supply a large volume of fundus images with DR grading and lesion-level annotations. For glaucoma research (Table \ref{adg-dataset}), resources such as ORIGA, REFUGE, and GAMMA combine structural information (e.g., optic disc imaging) with functional assessments to support detection and monitoring. Additional datasets (Table \ref{o-dataset}), including DRIVE for vessel segmentation and BIOMISA for multi-disease diagnosis, address broader ophthalmic tasks and enable wider research applications.

These datasets differ in scale, modality coverage and accessibility. They reflect the wide range of focus areas in ophthalmic AI research. By organizing these resources in a systematic way, this review helps researchers select suitable datasets, reduces the entry barrier and promotes fair comparisons and reproducibility across various tasks and modalities. This approach ultimately accelerates advancement and supports practical adoption in this fast-developing field.

\begin{table}[t]
\centering
\fontsize{7}{9}\selectfont
\caption{Publicly available datasets for AMD and DME research.}
\label{amd-dataset}
\begin{tabularx}{\linewidth}{%
  >{\RaggedRight\arraybackslash}m{1.8cm}  
| >{\centering\arraybackslash}m{0.7cm}    
  >{\centering\arraybackslash}m{1cm}    
  >{\centering\arraybackslash}m{1.5cm}    
  >{\centering\arraybackslash}m{0.7cm}    
}
\toprule
\textbf{Dataset} & \textbf{Modality} & \textbf{Images} & \textbf{Description} & \textbf{Access} \\
\midrule
Rabbani-I \cite{data1} & OCT & 4,142 & AMD, DME & Public \\
\midrule
Rabbani-II \cite{data2} & OCT & 19 subjects & Healthy subjects only & Public \\
\midrule
Duke-I \cite{data3} & OCT & 38,400 & AMD & Public \\
\midrule
Duke-II \cite{data4} & OCT & 10 subjects & DME & Public \\
\midrule
Duke-III \cite{data5} & OCT & 45 subjects & AMD, DME & Public \\
\midrule
AREDS \cite{data6} & CFP & 188,006 & AMD & Public \\
\midrule
AREDS2 \cite{data7} & CFP & 600,000+ & CFP only; FAF collected later & Public \\
\midrule
AV-WIDE \cite{data8} & UWF & 30 & AMD  & Public \\
\midrule
KORA \cite{data9} & CFP & 2,546 & AMD & Public \\
\midrule
OCTID \cite{data10} & OCT & 500 & AMD, DME  & Public \\
\midrule
iChallenge-AMD \cite{data11_icha} & CFP & 1,200 & AMD & Public \\
\bottomrule
\end{tabularx}
\label{tab:amd_datasets}
\end{table}

\begin{table}[t]
\centering
\fontsize{7}{9}\selectfont
\caption{Publicly available datasets for diabetic retinopathy (DR) research.}
\label{dr-dataset}
\begin{tabularx}{\linewidth}{%
  >{\RaggedRight\arraybackslash}m{2cm}  
| >{\centering\arraybackslash}m{0.7cm}    
  >{\centering\arraybackslash}m{0.7cm}    
  >{\centering\arraybackslash}m{1.5cm}    
  >{\centering\arraybackslash}m{0.7cm}    
}
\toprule
\textbf{Dataset} & \textbf{Modality} & \textbf{Images} & \textbf{Description} & \textbf{Access} \\
\midrule
DIARETDB \cite{data12} & CFP & 219 & DR & Public \\
\midrule
DeepDRiD \cite{data13} & CFP & 2,256 & DR & Public \\
\midrule
APTOS-2019 \cite{data14} & CFP & 3,662 & DR  & Public \\
\midrule
DDR \cite{data15} & CFP & 13,673 & DR  & Public \\
\midrule
Kaggle DR \cite{data16} & CFP & 88,702 & DR  & Public \\
\midrule
ROC \cite{data17} & CFP & 100 & DR & Public \\
\midrule
IDRiD \cite{data18} & CFP & 516 & DR, DME & Public \\
\midrule
ARIA \cite{data19} & CFP & 143 & DR, AMD  & Public \\
\midrule
Messidor \cite{data20} & CFP & 1,200 & DR, AMD & Public \\
\midrule
Messidor-2 \cite{data20}& CFP & 1,748 & DR  & Public \\
\midrule
Isfahan MISP \cite{data21_isf} & CFP, FFA & 70 & Paired CFP and FFA & Public \\
\bottomrule
\end{tabularx}
\label{tab:dr_datasets}
\end{table}

\begin{table}[t]
\centering
\fontsize{7}{9}\selectfont
\caption{Publicly available datasets for glaucoma research.}
\label{adg-dataset}
\begin{tabularx}{\linewidth}{%
  >{\RaggedRight\arraybackslash}m{2.2cm}  
| >{\centering\arraybackslash}m{0.7cm}    
  >{\centering\arraybackslash}m{0.7cm}    
  >{\centering\arraybackslash}m{1.3cm}    
  >{\centering\arraybackslash}m{0.7cm}    
}
\toprule
\textbf{Dataset} & \textbf{Modality} & \textbf{Images} & \textbf{Description} & \textbf{Access} \\
\midrule
LES-AV \cite{data22} & CFP & 22 & Glaucoma & Public \\
\midrule
ORIGA \cite{data23} & CFP & 650 & Glaucoma & Private \\
\midrule
RIM-ONE \cite{data24} & CFP (ONH) & 469 & Glaucoma & Public \\
\midrule
DRISHTI-GS \cite{data25_dris} & CFP & 101 & Glaucoma & Public \\
\midrule
REFUGE \cite{data26} & CFP & 1,200 & Glaucoma & Public \\
\midrule
Drions-DB \cite{data27} & CFP (ONH) & 110 & Glaucoma & Public \\
\midrule
RIGA \cite{data28} & CFP & 750 & Glaucoma & Public \\
\midrule
LAG \cite{data29} & CFP & 5,824 & Glaucoma & Public \\
\midrule
Rotterdam EyePACS AIROGS \cite{data30} & CFP & 113,893 & Glaucoma  & Public \\
\midrule
EyePACS-AIROGS-light \cite{data30} & CFP & 3,270 & Subset of Rotterdam EyePACS & Public \\
\midrule
EyePACS-AIROGS-light-v2 \cite{data30}& CFP & 4,770 & Extended subset & Public \\
\midrule
GAMMA \cite{Glau2023_GAMMA} & OCT (2D/3D) & 100 & Glaucoma  & Public \\
\bottomrule
\end{tabularx}
\label{tab:glaucoma_datasets}
\end{table}

\begin{table}[t]
\centering
\fontsize{7}{9}\selectfont
\caption{Publicly available datasets for other ophthalmic diseases and general research.}
\label{o-dataset}
\begin{tabularx}{\linewidth}{%
  >{\RaggedRight\arraybackslash}m{1.6cm}  
| >{\centering\arraybackslash}m{0.7cm}    
  >{\centering\arraybackslash}m{1cm}    
  >{\centering\arraybackslash}m{1.7cm}    
  >{\centering\arraybackslash}m{0.7cm}    
}
\toprule
\textbf{Dataset} & \textbf{Modality} & \textbf{Images} & \textbf{Description} & \textbf{Access} \\
\midrule
DRIVE \cite{data31_drive} & CFP & 40 & Vessel segmentation & Public \\
\midrule
CHASE-DB1 \cite{data32} & CFP & 28 & Vessel segmentation & Public \\
\midrule
HRF \cite{data33} & CFP & 45 & Vessel segmentation & Public \\
\midrule
STARE \cite{data34} & CFP & 20 & Vascular analysis. & Public \\
\midrule
MMR \cite{seg5} & CFP, OCT & 370 pairs & RAO & Private \\
\midrule
INSPIRE-stereo \cite{data35} & SFI & 30 pairs & 3D reconstruction of Glaucoma and ONH & Public \\
\midrule
BIOMISA \cite{data36} & CFP, OCT & 64 + 2,497 & Multi-eye disease diagnosis & Private \\
\midrule
Kermany et al. \cite{limit2}  & OCT & 108,312 & Multi-eye disease diagnosis & Public \\
\midrule
OCTA-500 \cite{data37} & OCTA & 500 & Multi-eye disease diagnosis & Public \\
\midrule
ROSE \cite{data38} & OCTA & 229 & Multi-disease diagnosis & Public \\
\bottomrule
\end{tabularx}
\label{tab:other_datasets}
\end{table}

\section{Future Discussion and Conclusion}
\subsection{Limitations and Challenges}
One major challenge facing current multimodal ophthalmic learning systems is the heterogeneity of data sources and annotation bottlenecks. In clinical reality, ophthalmic datasets are typically assembled from disparate imaging modalities each acquired using different devices, protocols, and resolutions. This results in substantial variability in image quality, illumination, field-of-view, and spatial scale, all of which pose significant obstacles for downstream model training and inference, making it difficult for deep learning models to generalize across cases. In cross-institutional scenarios, these discrepancies become even more pronounced due to different vendor-specific imaging pipelines \cite{limit1}.

In addition to inconsistencies in imaging, the scarcity of high-quality annotations creates a major bottleneck for model development. Expert-level labeling, especially pixel-level segmentation needed for lesion detection in datasets such as IDRiD or DDR, requires intensive labor, considerable time, and significant expense. This limitation affects not only the quantity but also the diversity of annotated data available for supervised learning. Moreover, manual annotation of multimodal images often demands distinct expertise, for example, retinal specialists for OCT and angiography experts for fluorescein fundus angiography \cite{limit2}, which further complicates the preparation of data. Although weakly supervised and self-supervised methods provide some mitigation, they frequently fall short in meeting the demands of clinical applications that require high precision.

Another persistent barrier to clinical integration lies in the limited interpretability of deep learning models, which undermines clinician trust and regulatory acceptance. Many task-specific fusion architectures, especially convolutional models, operate as opaque systems that provide diagnostic predictions without revealing the underlying decision rationale. Even with the advent of foundation models like RET-CLIP, which incorporate contrastive learning to align image and textual representations, the interpretability of their multi-head attention layers remains limited. Clinicians require visualization tools that not only highlight salient regions but also trace modality-specific reasoning in a clinically intuitive manner \cite{limit3}.

A further limitation lies in the modality gap and inefficiencies in feature fusion, which prevent current systems from fully exploiting the complementary advantages of different imaging types. Fundus photography effectively visualizes surface-level vascular changes, hemorrhages, and exudates, whereas OCT captures cross-sectional retinal morphology with high precision. Nevertheless, most existing fusion approaches, including early fusion that concatenates raw images or features and late fusion that combines modality-specific predictions, remain static and fail to achieve the necessary semantic alignment for meaningful integration of multi-scale and cross-domain features \cite{limit4, limit5}. As a result, performance in complex tasks is often suboptimal. Furthermore, current fusion methods generally consider each modality equally informative across patients and disease conditions, overlooking the context-dependent relevance of specific modalities. This oversimplification restricts the diagnostic robustness.

Finally, limited generalization to demographically and geographically diverse populations poses a major threat to the clinical deployment of multimodal ophthalmic AI. Most existing datasets originate from single-region hospitals or academic centers and are heavily skewed toward specific ethnic or socioeconomic groups. For instance, the REFUGE dataset is dominated by Chinese patient populations, which raises concerns about algorithmic bias when deployed in Western or African settings. Similarly, the GAMMA dataset, largely based on Western cohorts, has shown decreased performance when tested on populations from low-resource settings where disease presentation and image quality may differ. This lack of population diversity not only limits external validity but also risks reinforcing healthcare inequities \cite{limit6}. In addition, existing benchmarks rarely stratify performance by demographic attributes such as age, sex, or comorbidities, factors that are known to influence both imaging biomarkers and disease progression. The lack of demographic stratification further obscures potential model limitations, making it more difficult to identify whether diagnostic tools perform consistently across diverse patient populations.

Overall, these limitations underscore the urgent need for standardized data acquisition protocols, scalable annotation strategies, interpretable model design, and demographically inclusive evaluation frameworks to ensure that multimodal AI systems are not only accurate, but also trustworthy and equitable across global clinical contexts.

\subsection{Future Research Directions}
Ultra-widefield (UWF) imaging provides a fertile ground for advancing multimodal deep learning in ophthalmology by enabling the integration of structural, vascular, and clinical modalities. UWF captures up to 200° of the retina in a single image, offering access to peripheral retinal regions typically excluded in conventional fundus photography (30°–60°), thus facilitating earlier and more holistic disease characterization. Recent multimodal frameworks such as UWF-CKDS have combined UWF-based vessel morphology (e.g., fractal dimension, tortuosity) with clinical biomarkers (e.g., eGFR, diabetes history), achieving robust performance in systemic disease prediction, with AUCs up to 0.86 for chronic kidney disease screening \cite{uwf1}. Other studies such as Multi-disease UWF-DL and DeepUWF-plus \cite{uwf2, uwf6} incorporate UWF imaging with multi-label learning or hierarchical classification to simultaneously detect multiple coexisting retinal conditions, leveraging both central and peripheral retinal cues. Additionally, combining UWF with emerging modalities like fundus autofluorescence (FAF) and optical coherence tomography angiography (OCTA) has proven valuable for estimating visual function and characterizing metabolic or microvascular changes in retinal diseases such as retinitis pigmentosa and diabetic retinopathy \cite{uwf3, uwf4}. Future directions may focus on developing efficient multimodal fusion strategies, lightweight deployment-ready architectures for use in primary care, and enhanced modeling of peripheral biomarkers to improve generalizability in real-world, longitudinal, and population-scale ophthalmic studies \cite{uwf1, uwf2, uwf6, uwf7}.

Future ophthalmic AI systems must do more than achieve high diagnostic accuracy. They should incorporate explainable and step-by-step reasoning capabilities through multimodal learning mechanisms. A key area for future development is large model frameworks based on reinforcement learning (RL), which aim to emulate clinician decision-making processes and deepen the understanding of models of complex clinical information. In this regard, DeepSeek-R1 \cite{fut1} makes a foundational contribution. It departs from traditional purely supervised training approaches by using a hybrid approach that combines a cold-start phase with reinforcement learning to optimize reasoning trajectories. DeepSeek-R1-Zero further advances this concept by completely omitting supervised pre-training and relying exclusively on reinforcement learning to develop high-quality long-chain reasoning capabilities. This feature is particularly valuable in medical fields such as ophthalmology, where annotated data are often scarce. Such models can generate structured and logically coherent diagnostic paths, providing a strong foundation for explainable clinical decision support.

Building on this foundation, Vision-R1 and VLM-R1 \cite{fut2, fut3} apply reinforcement learning driven reasoning mechanisms within multimodal deep learning frameworks. This approach fits well with ophthalmology because it allows integration of multiple inputs, including ultra-widefield imaging, fundus photography, and fluorescein angiography, in order to achieve precise lesion detection and pathology localization. These works collectively show that reinforcement learning not only enhances reasoning abilities but also supports the development of stable and transferable strategies across different multimodal domains. Examples of adapting such multimodal reasoning models to medical applications include Med-R1 and MedVLM-R1 \cite{fut4, fut5}. Their training regime improves model generalizability and demonstrates the effectiveness of reinforcement learning in building explainable models under conditions of limited annotated data. For ophthalmology, this indicates a promising direction toward developing compact yet intelligent multimodal diagnostic systems capable of performing well in real-world clinical environments characterized by scarce data and diverse imaging modalities.

\subsection{Conclusion}
This survey reviewed recent advances in multimodal deep learning for ophthalmic diagnostics, focusing on task-specific models and large-scale foundation models. The integration of multiple imaging modalities with clinical and textual data significantly improves diagnostic accuracy and generalization in diseases such as diabetic retinopathy, glaucoma, and age-related macular degeneration. Key methodological innovations include self-supervised learning, attention-based fusion, and contrastive alignment, while persistent challenges involve data heterogeneity, annotation scarcity, interpretability, and population diversity. Future research should explore ultra-widefield imaging for broader retinal assessment, reinforcement learning-driven multimodal reasoning to mimic clinical decision-making, and domain adaptation techniques to enhance robustness. Emphasizing explainability will facilitate clinical adoption and regulatory compliance.




\bibliography{references}
\end{document}